\documentclass[prd,showkeys,superscriptaddress,nofootinbib,floatfix,preprint,12pt]{revtex4-2}

\usepackage[colorlinks=true,
breaklinks=true,
urlcolor=blue,
citecolor=blue]{hyperref}

\usepackage{graphicx} 
\usepackage{slashed,color,amsmath,amssymb,mathrsfs}
\usepackage{graphicx}
\usepackage{subfigure}
\usepackage{booktabs}
\usepackage{hyperref}
\usepackage{longtable}
\usepackage{array}
\usepackage{enumerate}
\usepackage{enumitem}

\usepackage{indentfirst}
\usepackage{verbatim}
\usepackage{makecell}
\usepackage{multirow}
\usepackage{bm}
\usepackage[normalem]{ulem}

\begin{document}
\title{Neutrinoless double-beta decay of the $\Delta^-$ resonance}

\author{Li-Ping~He}
\email{heliping@csu.edu.cn}
\affiliation{School of Physics, Central South University, Changsha 410083, China}
\affiliation{Helmholtz-Institut f\"ur Strahlen- und Kernphysik and Bethe Center for Theoretical Physics, Universit\"at Bonn, D-53115 Bonn, Germany}

\author{Feng-Kun~Guo}
\email{fkguo@itp.ac.cn}
\affiliation{Institute of Theoretical Physics, Chinese Academy of Sciences, Beijing 100190, China}
\affiliation{School of Physical Sciences, University of Chinese Academy of Sciences,\\ Beijing 100049, China}
\affiliation{Southern Center for Nuclear-Science Theory (SCNT), Institute of Modern Physics, Chinese Academy of Sciences, Huizhou 516000, China}

\author{Ulf-G. Mei\ss ner}
\email{meissner@hiskp.uni-bonn.de}
\affiliation{Helmholtz-Institut f\"ur Strahlen- und Kernphysik and Bethe Center for Theoretical Physics, Universit\"at Bonn, D-53115 Bonn, Germany}
\affiliation{Institute~for~Advanced~Simulation (IAS-4), Forschungszentrum~J\"{u}lich, D-52425~J\"{u}lich,~Germany}
\affiliation{Peng Huanwu Collaborative Center for Research and Education, International Institute for Interdisciplinary and Frontiers, Beihang University, Beijing 100191, China}

\author{De-Liang~Yao}
\email{yaodeliang@hnu.edu.cn}
\affiliation{School of Physics and Electronics, Hunan University, Changsha 410082, China}
\affiliation{Hunan Provincial Key Laboratory of High-Energy Scale Physics and Applications, Hunan University, Changsha 410082, China}

\author{Xiao-Yu~Zhang}
\affiliation{Institute of Theoretical Physics, Chinese Academy of Sciences, Beijing 100190, China}

\author{Zhen-Hua~Zhang}
\affiliation{Center for High Energy Physics, Peking University, Beijing 100871, China}
\affiliation{School of Physics, Peking University, Beijing, 100871, China}

\date{\today}

\begin{abstract}

    The subprocess $nn\to ppe^-e^-$ is a key ingredient in the interpretation of nuclear neutrinoless double-beta decay. 
    Intermediate $\Delta$ resonances may provide additional enhancements to this transition. We take a first step toward a $\Delta$-full description of $nn\to ppe^-e^-$ by investigating the neutrinoless double-beta decay $\Delta^- \to p e^-e^-$ in the framework of chiral effective field theory. 
    We systematically derive the long-range contribution from light-Majorana-neutrino exchange through loop diagrams and incorporate the short-range part through counterterms required by renormalization. We predict the pion-mass dependence of the decay amplitude in the kinematic configuration with collinear electrons. 
    Furthermore, to facilitate lattice-QCD matching, we calculate the decay amplitude in the degenerate $\Delta$-nucleon mass limit and provide the corresponding long-range prediction. 

\end{abstract}

\maketitle

\newpage

\section{Introduction}\label{sec:1}

In the Standard Model (SM) of particle physics, neutrinos are described as massless fundamental particles. 
However, the observation of neutrino oscillations~\cite{Super-Kamiokande:1998kpq,SNO:2002tuh,KamLAND:2002uet} provided conclusive evidence that neutrinos must have non-zero mass, showing that the SM must be extended. 
This can be done simply by adding Yukawas for the
neutrinos or invoking physics beyond the SM.
However, neutrino oscillations are insensitive to the absolute neutrino mass scale and cannot determine whether they are Dirac or Majorana particles. 
To address these fundamental questions, complementary approaches from laboratory experiments and cosmological observations are essential. 
The most recent direct tritium-$\beta$-decay measurement by the KATRIN Collaboration gives the strongest laboratory upper limit on the effective electron-neutrino mass, $m_\nu < 0.45$~eV at 90\% confidence level~\cite{KATRIN:2024cdt}.  
Model-dependent cosmological analyses place even tighter upper limits: 0.12~eV from the Planck Collaboration~\cite{Planck:2018vyg} and 0.072~eV from the DESI Collaboration~\cite{DESI:2024mwx}. 
Neutrinoless double-beta ($0\nu\beta\beta$) decay, in which two electrons are emitted without their corresponding antineutrinos, offers a third avenue to investigate neutrino mass~\cite{Schechter:1981bd}. 
The most stringent current upper limit on the effective Majorana neutrino mass comes from the KamLAND-Zen Collaboration, which finds a range of 28--122~meV from the $0\nu\beta\beta$ decay of $^{136}$Xe~\cite{KamLAND-Zen:2024eml}. 

The observation of $0\nu\beta\beta$ decay not only provides a way to determine the effective neutrino mass but also demonstrates that neutrinos are Majorana particles and the total lepton number $L$ is violated by two units, i.e., $\Delta L=2$. This process may relate to the matter-antimatter asymmetry in the universe via ``leptogenesis''~\cite{Davidson:2008bu} and serve as a probe of other beyond-SM theories. 
The significance of $0\nu\beta\beta$ decay has motivated numerous experimental searches in both nuclear  \cite{LEGEND:2025jwu,CUPID:2025faj,KamLAND-Zen:2024eml,PandaX:2024fed,AMoRE:2024loj,NEXT:2023daz,CDEX:2023owy,Majorana:2022udl,CUPID:2022puj,XENON:2022evz,CUORE:2021mvw,GERDA:2020xhi} 
and hadronic physics \cite{BESIII:2025ylz,BESIII:2025qjn,BESIII:2025gsy,BESIII:2024ziy,BESIII:2020iwk,BESIII:2019oef,LHCb:2020car,LHCb:2012pcm,HyperCP:2005sby}. 
However, the process is highly suppressed because the matrix element is proportional to the Fermi coupling constant squared ($G_F^2$) and the tiny effective Majorana neutrino mass $m_{\beta\beta}$, or to couplings of the same order. 
The interpretation of nuclear $0\nu\beta\beta$ decay relies on complicated nuclear matrix elements. 
Current theoretical calculations can differ by about one order of magnitude in their predicted half-lives for nuclear $0\nu\beta\beta$ decay~\cite{Brase:2021uny,Agostini:2022zub,Yao:2021wst,Dolinski:2019nrj}. 
Nevertheless, the matrix elements of $0\nu\beta\beta$ decay for nucleons and hadrons can be systematically computed with controlled accuracy using chiral effective field theory ($\chi$EFT)~\cite{Hammer:2019poc,Machleidt:2011zz,Epelbaum:2008ga}.  

The $0\nu\beta\beta$ transition of two neutrons into two protons, $nn\to pp e^-e^-$, constitutes an elementary subprocess of various nuclear $0\nu\beta\beta$ decays, making it crucial for interpreting experimental searches. 
Its corresponding matrix element has been studied up to next-to-next-to-leading order (NNLO) in $\chi$EFT~\cite{Cirigliano:2017tvr,Pastore:2017ofx,Cirigliano:2018yza,Cirigliano:2018hja,Wang:2018htk,Yang:2023ynp,Yang:2025tdw}, where the long-range and short-range contributions can be identified systematically. 
The long-range contribution can be mediated by light Majorana-neutrino exchange~\cite{Bilenky:2014uka}. In the SM effective-field-theory description, this mechanism originates from the well-known dimension-five Weinberg operator~\cite{Weinberg:1979sa}, which provides the leading low-energy manifestation of lepton number violation (LNV) and generates the light Majorana mass of light neutrinos when the new-physics scale $\Lambda_{\rm LNV}$ is much larger than the electroweak scale $\Lambda_{\rm EW}$. 
The short-range contribution accounts for higher-energy-scale physics, described by various beyond-SM models, and may not involve Majorana neutrinos directly.
In the previous works, the $\Delta(1232)$ resonance, denoted by $\Delta$ for short hereafter, was not included explicitly.
Because the $\Delta$ couples strongly to pions and nucleons, many observables are described more accurately in a $\Delta$-full theory than in a $\Delta$-less one; see, e.g., Refs.~\cite{Hemmert:1997ye,Yao:2016vbz}. Furthermore, the relevant low-energy constants (LECs) become more natural~\cite{Bernard:2007zu,Krebs:2007rh,Epelbaum:2008ga,Machleidt:2011zz,Siemens:2016jwj,Yao:2016vbz,Ekstrom:2017koy}. In this regard, it is essential to derive the matrix element of the $0\nu\beta\beta$ transition $nn\to pp$ in a $\Delta$-full theory. 
Recently, the three-nucleon neutrinoless double-$\beta$ decay potential in a $\Delta$-full $\chi$EFT was studied in Ref.~\cite{Chambers-Wall:2025zfw}. 

\begin{figure}[t]
\includegraphics*[width=0.25\linewidth]{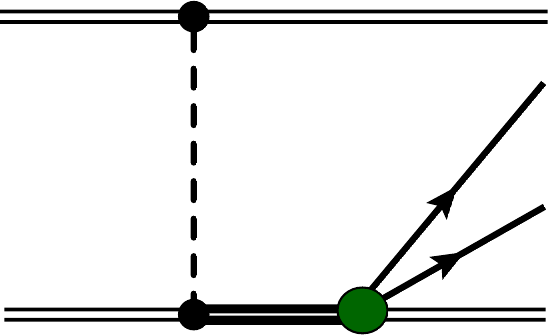}
\caption{
Illustration of intermediate $\Delta$ contribution to the $0\nu\beta\beta$ process $nn \to pp e^-e^-$.
Electrons are denoted by solid lines with arrows and pions by dashed lines. Nucleons and $\Delta$ resonances are represented by thin and thick double lines, respectively. The $\Delta L=2$ vertex of $\Delta^-\to pe^-e^-$ is indicated by a dark green blob.
}
\label{fig:Feyn-nn-Delta-ppee}
\end{figure}

In this work, we take a first step toward a $\Delta$-full theory description of $nn\to ppe^-e^-$ by calculating the $0\nu\beta\beta$ matrix element of the $\Delta^-\to pe^-e^-$ decay in the framework of $\chi$EFT. It contributes to the $nn\to ppe^-e^-$ process through pion and $\Delta$ exchanges between nucleons, as illustrated in Fig.~\ref{fig:Feyn-nn-Delta-ppee}. The intermediate $\Delta$ exchange in Fig.~\ref{fig:Feyn-nn-Delta-ppee} begins to contribute at NNLO in the Weinberg power counting~\cite{Weinberg:1990rz,Weinberg:1991um}, i.e., $\mathcal{O}(p^2)$, 
the same order as those calculated in Ref.~\cite{Cirigliano:2017tvr}. The dark green blob includes long- and short-range contributions. The long-range contribution arises from the exchange of light Majorana neutrinos through one-loop diagrams. 
Some of these loop diagrams develop threshold cusps and triangle singularities, which may significantly enhance the long-range contribution and make it more important than other NNLO contributions to nuclear $0\nu\beta\beta$ decay. 
The short-range contribution arises from counterterm operators that are required to renormalize the ultraviolet divergences from loop diagrams. We construct the pertinent contact $\Delta^-\to pe^-e^-$ operators, invariant under chiral transformations and starting at $\mathcal{O}(p^3)$ in the naive chiral expansion. The counterterms introduce extra unknown LECs, which can be determined by Lattice Quantum Chromodynamics (LQCD) simulations in the future.
At sufficiently large pion masses, the $\Delta$ resonance becomes stable against strong decays and can therefore be treated as an asymptotic state.
A particularly suitable setup for LQCD studies is the mass-degenerate limit of the $\Delta$ and the proton. In this limit, the matrix element is purely real and becomes analytically manageable, facilitating a direct comparison between LQCD results and $\chi$EFT predictions.

The remainder of this paper is organized as follows. 
In Sec.~\ref{sec:framework}, we describe the light-Majorana-neutrino exchange mechanism and the $\chi$EFT framework employed in our calculation.  
In Sec.~\ref{sec:Deltatopee}, we present the calculation of the $\Delta^-\to pe^-e^-$ matrix element and construct the relevant contact interactions. 
The pion mass dependence of the decay amplitude, for both the physical $\Delta$ mass and the degenerate $\Delta$-proton mass limit, is discussed in Sec.~\ref{sec:pionmass}. 
We conclude with a summary in Sec.~\ref{sec:sum}.

\section{Effective field theory framework}
\label{sec:framework}
In this section, we describe the LNV operator for the $0\nu\beta\beta$ decay and introduce the chiral effective Lagrangians relevant to our calculation.

\subsection{The LNV effective Lagrangian
}
\label{subsec:neutrino}
There is fundamental interest in understanding the origin of the small masses of neutrinos. 
If they are generated by the Brout-Englert-Higgs mechanism, the dimensionless neutrino Yukawa coupling constants should be many orders of magnitude smaller than the coupling constants of other leptons and quarks. It is highly unlikely that neutrino masses are of the SM origin. 
It is straightforward to extend the SM by adding right-handed neutrinos through the seesaw mechanism~\cite{Minkowski:1977sc,Gell-Mann:1979vob,Yanagida:1979as,Schechter:1980gr,Mohapatra:1979ia}.  

Nevertheless, a more economical approach, referred to as the minimal scenario, is to introduce only left-handed Majorana mass terms, which is realized by the Weinberg effective Lagrangian \cite{Weinberg:1979sa},
\begin{equation}
\mathcal{L}^{\mathrm{eff}} = 
- \frac{1}{\Lambda_\mathrm{LNV}} \big(\bar\psi_{L}\tilde{H}\big) Y C \big(\bar \psi_{L}\tilde{H}\big)^T + \text{H.c.}\ ,
\end{equation}
where $C$ denotes the charge conjugation operator, $\Lambda_\mathrm{LNV}$ is an energy scale much larger than the Higgs vacuum expectation value $v \sim 246$~GeV, and $Y$ is a $3\times 3$ symmetric matrix in flavor space not constrained by the $SU_L(2)\times U_Y(1)$ symmetry. 
This is the leading low-energy manifestation of LNV if $\Lambda_\mathrm{LNV}$ is much larger than the electroweak scale.
After the spontaneous symmetry breaking,
\begin{equation}
\bar\psi_{L}\tilde{H} = \frac{v+h}{\sqrt{2}} \bar \nu_L\ ,
\end{equation}
where the left-handed neutrino field is defined as $\bar \nu_L = (\bar \nu_{e,L}, \bar\nu_{\mu,L}, \bar \nu_{\tau,L})$, one obtains the following Majorana neutrino mass term
\begin{equation}
\mathcal{L}^{\mathrm{eff}}_\mathrm{M} =
- \frac{1}{2} \frac{v^2}{\Lambda_\mathrm{LNV}} \bar \nu_L Y C\bar\nu_L^T + \text{H.c.}\ .
\label{eq:L-mass-flavor}
\end{equation}
It describes the transition between a left-handed neutrino and its charge conjugate, which is itself. 
The mass term involves the annihilation or creation of two neutrinos, and therefore a violation of lepton number by two units, $\Delta L =2$. 
Such LNV is suppressed by the ratio $v/\Lambda_{\mathrm{LNV}}$. 

The symmetric matrix $Y$ can be diagonalized by a unitary matrix $U$ via $Y=U\mathcal{Y}U^T$, where $\mathcal{Y}_{ij} = y_i \delta_{ij}$. The matrix $U$ is called the Pontecorvo-Maki-Nakagawa-Sakata (PMNS) mixing matrix, similar to the Cabibbo–Kobayashi–Maskawa (CKM) matrix in the quark sector.
The flavor field $\nu_{L}$ is related to the mass eigenstate neutrino field $\nu_{L}^\mathrm{m}$ by $ \nu_{L}  = U  \nu_{L}^\mathrm{m}$.
The mass term in Eq.~\eqref{eq:L-mass-flavor} can then be written in the standard form of a Majorana neutrino mass term
\begin{equation}
\mathcal{L}^{\mathrm{eff}}_\mathrm{M} =
- \frac{1}{2} \sum_{i=1}^3 \bar \nu_i   m_i \nu_i, 
\label{eq:L-mass}
\end{equation}
where $\nu_i = \nu_{i,L}^\mathrm{m} + C(\bar\nu_{i,L}^\mathrm{m})^T$, and $m_i = v^2 y_i/\Lambda_\mathrm{LNV}~(i=1,2,3)$ is the mass of the neutrino mass eigenstate. In consequence, the Majorana mass term in Eq.~\eqref{eq:L-mass-flavor} can be recast to
\begin{equation}
    \mathcal{L}_{\beta\beta} = -\frac12 m_{\beta\beta} \left(\nu_L^TC\nu_{L}^{} + \bar\nu_L^{} C \bar\nu_L^T\right),\quad m_{\beta\beta} = \sum_{i=1}^3U_{1i}^2 m_i\ ,
    \label{eq:Lag-mbetabeta}
\end{equation}
where $m_{\beta\beta}$ is usually called the effective Majorana neutrino mass.

\subsection{Chiral effective field Lagrangians}
\label{subsec:baryonEFT}
The spontaneous breaking of chiral symmetry of quantum chromodynamics (QCD) provides strong constraints for the low-energy interactions among nucleons and pions. The Goldstone nature of the pions requires derivative interactions. 
Furtheremore, chiral symmetry is explicitly broken by the light quark masses $m_q$, which are smaller than the nonperturbative QCD scale $\Lambda_{\rm QCD}$. 
In the low-energy region, one can construct a systematic expansion in powers of $p/\Lambda_\chi$, where $p$ collectively denotes small quantities involved in a given process, including the small momenta and pseudo-Nambu-Goldstone boson masses, and $\Lambda_\chi$ ($\sim 1$~GeV) is the chiral symmetry breaking scale. 

The leading-order (LO) effective Lagrangian describing interactions among pions, leptons and nucleons reads~\cite{Gasser:1983yg,Fettes:2000gb} (for a detailed 
exposition, see~\cite{Meissner:2022cbi})
\begin{subequations}
\begin{eqnarray}
\mathcal{L}_{\pi\pi}^{(2)}  &=&  \frac{F^2}{4} \text{Tr}\left[ u_\mu u^\mu + u^\dagger \chi u^\dagger + u\chi^\dagger u \right]\ ,
\label{eq:Lpion}
 \\ 
\mathcal{L}_{\pi N}^{(1)} 
&=& 
\bar N \left[ i (\slashed\partial + \slashed\Gamma)  - m_N  + \tfrac12 g_A\, \slashed u\gamma_5 \right]N\ ,
\label{eq:LpiN}
\end{eqnarray}
\label{eq:Lagrangian}%
\end{subequations}
where the superscripts in brackets indicate the respective chiral orders, $F$ is the pion decay constant in the chiral limit, and $\chi$ is a diagonal mass matrix associated with the charged pion mass by $\chi = M_{\pi^\pm}^2 \mathbf{1}$. Here, $m_N$ and $g_A$ are the mass and axial coupling constant of the nucleon in the chiral limit, respectively. 
We set $g_A = 1.2753(13)$~\cite{ParticleDataGroup:2024cfk} in our numerical computation below. 
The pion fields are non-linearly parametrized as $u = \exp[i\bm \pi \cdot \bm \tau/(2F)]$, with $\bm \tau = (\tau^1, \tau^2, \tau^3)$ being the Pauli matrices in isospin space.
It transforms as $u\to LuK^\dagger(\pi) = K(\pi) uR^\dagger$ under $SU(2)_L \times SU(2)_R$ chiral group \cite{Coleman:1969sm,Callan:1969sn}. 
The nucleon doublet $N = (p,n)^T$ transforms as $N\to K(\pi) N$. The chiral vielbein (axial vector) $u_\mu$ and chiral connection $\Gamma_\mu$ are defined by
\begin{subequations}
    \begin{eqnarray}
        u_\mu 
         &=&
         -i \left( u^\dagger \partial_\mu u  -u \partial_\mu u^\dagger\right) - u^\dagger \ell_\mu u\ ,  \\
         \label{eq:def-u_mu}
        \Gamma_\mu 
        &=&  
        \frac{1}{2}\left(u^\dagger \partial_\mu u+ u\partial_\mu u^\dagger\right) -\frac{i}{2}u^\dagger \ell_\mu u\ ,
        \label{eq:def-Gamma_mu}
    \end{eqnarray}
\end{subequations}
where the coupling to the left-handed leptons is incorporated by setting
\begin{align}
  \ell_\mu = -2\sqrt{2} G_F V_{ud} \tau^+ \bar e_L \gamma_\mu \nu_L +\text{H.c.}  \ ,
\end{align}
with $G_F$ the Fermi constant and $V_{ud}$ the pertinent CKM matrix element.

The $\Delta$ resonance has isospin 3/2 and spin 3/2. 
We denote the four $\Delta$ isospin states concisely as $\Delta_\mu = \left( \Delta^{++}_\mu, \Delta^{+}_\mu, \Delta^{0}_\mu, \Delta^{-}_\mu\right)^T$.
Each component is a spin-$3/2$ field that can be described by the Rarita-Schwinger formalism \cite{Rarita:1941mf}. 
It has the mixed transformation properties of a Dirac four-component of wavefunction and of a four-vector field (and of asymmetric tensor of rank 4). 
The four $\Delta$ isospin states can be regarded as the isospin-$3/2$ component of the tensor product $1 \otimes 1/2$ in the isovector-isospinor formalism.
They can be conveniently written in the isospurion notation $\Psi^i_\mu(x)$~\cite{Tang:1996sq}:
\begin{align}
\Psi^1_\mu 
         &=
         \frac{1}{\sqrt{2}}\left( 
         \begin{array}{cc}
           \frac{1}{\sqrt{3}} \Delta^0_\mu - \Delta^{++}_\mu \\
            \Delta^-_\mu - \frac{1}{\sqrt{3}} \Delta^{+}_\mu 
         \end{array}\right )\ ,\notag \\
\Psi^2_\mu 
         &=
         -\frac{i}{\sqrt{2}}\left( 
         \begin{array}{cc}
           \frac{1}{\sqrt{3}} \Delta^0_\mu + \Delta^{++}_\mu \\
            \Delta^-_\mu + \frac{1}{\sqrt{3}} \Delta^{+}_\mu 
         \end{array}\right )\ ,  \notag \\
\Psi^3_\mu 
         &=
         \sqrt{\frac{2}{3}}\left( 
         \begin{array}{cc}
          \Delta^+_\mu\\
          \Delta^0_\mu
         \end{array}\right )\ ,
         \label{eq:def-psi}
\end{align}
where $\Psi_\mu^i(x)$ is an isovector-isospinor field: for each $i\in\{1,2,3\}$, $\Psi_\mu^i$ is an isospin doublet. The above representation can be obtained by applying the isospin-$3/2$ projector
\begin{equation}
P_{3/2}=\sum_{M=-3 / 2}^{3 / 2}\left|3/2, M\right\rangle\left\langle 3/2, M\right|
\end{equation}
to each of the six tensor products $|1,m\rangle\otimes|\tfrac12,r\rangle$ ($m=-1,0,1$ and $r=-1/2,1/2$), and organizing the resulting isospin-$3/2$ $\Delta$ states in the three-dimensional Cartesian basis. 
The other two isospin-1/2 states are automatically removed by the projection operator, leading to the constraint $\tau^i \Psi^i_\mu(x) = 0$.

In the isospurion notation, the LO Lagrangian for interactions among $\Delta$, nucleons, pions and leptons is given by~\cite{Hemmert:1997ye,Scherer:2012xha}
\begin{equation}
    \begin{split}
    \mathcal{L}_{\pi \Delta}^{(1)}
    =&
    -\bar{\Psi}_\mu^i \bigg[
    (i\slashed{D}^{ij}-m_\Delta\delta^{ij})g^{\mu\nu} + i A (\gamma^\mu D^{\nu,ij}+\gamma^\nu D^{\mu,ij})\\
    & +\frac{i}{2}(3A^2+2A+1)\gamma^\mu \slashed{D}^{ij}\gamma^\nu +m_\Delta (3A^2+3A+1)\gamma^\mu\gamma^\nu \delta^{ij} \\
    & +\frac{g_1}{2}\slashed{u}^{ij}\gamma_5g^{\mu\nu}+\frac{g_2}{2}(\gamma^\mu u^{\nu,ij}+ u^{\mu,ij} \gamma^\nu)\gamma_5+\frac{g_3}{2}\gamma^\mu\slashed{u}^{ij}\gamma_5\gamma^\nu\bigg]\Psi_\nu^j\ , 
    \end{split}
    \label{eq:Lag-piDelta}
\end{equation}
where $A~(A\neq -1/2)$ is an arbitrary real parameter, $m_\Delta$ is the mass of the $\Delta$ resonance in the chiral limit, and $u_\mu^{ij} = u_\mu \delta^{ij}$.
The covariant derivative acting on the $\Delta$ field is defined by 
\begin{equation}
     \mathcal{D}_{\mu}^{i j} =
        \partial_{\mu} \delta^{i j} -2 i \epsilon^{i j k} \Gamma_{\mu, k}+\delta^{i j} \Gamma_{\mu}\ ,
\end{equation}
with the chiral connection $\Gamma_\mu$ given by Eq.~\eqref{eq:def-Gamma_mu} and $\Gamma_\mu = \Gamma_{\mu,k} \tau_k$.
The last two terms in Eq.~\eqref{eq:Lag-piDelta} describe off-shell components of the $\Delta$ fields. 
It has been shown that they are redundant~\cite{Tang:1996sq,Pascalutsa:2000kd,Krebs:2009bf}, since their contribution can be absorbed into other parameters in the Lagrangian.
Hence, we take the off-shell parameters $g_2 = g_3 =0$ and $g_1 =-1.21 $~\cite{Yao:2016vbz}.

The lowest order Lagrangian for $\Delta$-nucleon transition can be expressed as~\cite{Hemmert:1997ye,Scherer:2012xha}
\begin{equation}
        \mathcal{L}_{\pi N \Delta}^{(1)} 
        =h \bar\Psi_\mu^i \Theta^{\mu\nu}(z) \omega_{\nu,i} N +{\rm H.c.},
        \label{eq:Lag-piNDelta}
\end{equation}
where $\omega_{\nu,i} = (1/2) \text{Tr}[\tau_i u_\nu] $, $h$ is the coupling constant, and $\Theta^{\mu\nu}(z) = g^{\mu\nu} + z \gamma^\mu \gamma^\nu$ with $z$ being another off-shell parameter. 
It has been shown that $z$ can be absorbed by shifting other LECs and therefore is arbitrary~\cite{Tang:1996sq,Pascalutsa:2000kd,Krebs:2009bf}. 
We set $z=0$ in what follows for convenience.
A fit to the decay width of $\Delta$ resonance leads to $h=1.05$~\cite{Hemmert:1997ye}. We treat the $\Delta$ resonance in the so-called small-scale-expansion (SSE) scheme~\cite{Hemmert:1996xg,Hemmert:1997ye}, where the mass difference $\delta=m_\Delta-m_N$ is counted as of order $\mathcal{O}(p)$.

\section{Decay amplitude of $\Delta^-\to p e^- e^-$}
\label{sec:Deltatopee}

In this section, we calculate the $0\nu\beta\beta$ decay $\Delta^-(p_1) \to p(p_2) e^-(q_1)e^-(q_2)$, where $p_1$, $p_2$, $q_1$ and $q_2$ denote the four-momenta of the $\Delta$, proton, and electrons, respectively.
The interaction vertices relevant to this process are shown explicitly in Appendix~\ref{App:Lagrangians}. 
Three Mandelstam variables are defined as
\begin{subequations}
\begin{eqnarray}    
s&=& (q_1+q_2)^2=(p_1-p_2)^2\ ,\\ 
t&=&(p_1-q_1)^2 =(q_2+p_2)^2\ ,\\ 
u&=&(p_1-q_2)^2=(q_1+p_2)^2\ ,
\end{eqnarray}
\label{eq:Mandelstam}
\end{subequations}
satisfying the constraint $s+t+u = m_\Delta^2 + m_N^2$ in the massless electron limit.

\subsection{Lorentz structure of the amplitude}
\label{subsec:Amp-Loop}
The decay amplitude can be generically expressed as
\begin{equation}
       \mathcal{M} = G\,
        \bar{u}_N\!\left(p_2\right)  \left[T_{S}^{\alpha} \bar{u}_L(q_1) C\bar{u}_L^T(q_2) - i T_{A}^{\alpha, \mu \nu}\bar{u}_L(q_1) \sigma_{\mu \nu} C\bar{u}_L^T(q_2)  \right] u_\alpha\!\left(p_1\right), 
        \label{eq:def-Amp}
\end{equation}
where the overall factor $G\equiv (2\sqrt{2}G_F V_{ud})^2 m_{\beta\beta} h$ and $\sigma_{\mu\nu} = i[\gamma_\mu, 
\gamma_\nu]/2$. Here, the symmetric and antisymmetric (under interchange of Lorentz indices $\mu\leftrightarrow\nu$) hadronic amplitudes read
\begin{subequations}
\begin{eqnarray}
    T_S^\alpha &=& (T_t^{\alpha,\mu\nu} + T_u^{\alpha,\mu\nu}) g_{\mu\nu}\ , 
    \label{eq:TS}\\ 
    T_{A}^{\alpha, \mu \nu} &=& T_t^{\alpha,\mu\nu} - T_u^{\alpha,\mu\nu},
    \label{eq:TA}
\end{eqnarray}
\end{subequations} 
where $T_t^{\alpha,\mu\nu}$ and $T_u^{\alpha,\mu\nu}$ denote the components arising from the $t$-channel and $u$-channel contributions.
They are related to each other by the interchange of the momenta $q_1$ and $q_2$, i.e.,
$T_t^{\alpha,\mu\nu}\leftrightarrow T_u^{\alpha,\mu\nu}$. 
Therefore, the antisymmetric hadronic amplitude $T_{A}^{\alpha, \mu \nu}$ in Eq.~\eqref{eq:def-Amp} vanishes if $q_1 = q_2$, since the $t$-channel and $u$-channel contributions cancel. 
In this work, we focus on the symmetric hadronic amplitude $T_S^\alpha$. 
It can be further decomposed into a linear combination of eight Lorentz structures as
\begin{equation}
       T_S^\alpha = T_1 q_1^\alpha + T_2 q_2^\alpha + T_3 q_1^\alpha\slashed{q}_2 + T_4 q_2^\alpha\slashed{q}_1+
      T_1^\prime q_1^\alpha\gamma_5 + T_2^\prime q_2^\alpha\gamma_5 + T_3^\prime q_1^\alpha\slashed{q}_2\gamma_5 + T_4^\prime q_2^\alpha\slashed{q}_1\gamma_5\ ,
        \label{eq:def-Amp-S}
\end{equation}
where the coefficients $T_i^{(\prime)} (i = 1, 2, 3, 4)$ are structure functions of the Mandelstam variables $s, t$, and $u$.  
Due to the symmetry of the two indistinguishable outgoing electrons, $T_1^{(\prime)}$ and $T_2^{(\prime)}$ are the same under the interchange $t\leftrightarrow u$. 
Similarly, the same symmetry applies to $T_3^{(\prime)}$ and $T_4^{(\prime)}$. 
Note that the equations of motions (EOMs) of the $\Delta$ resonance and the nucleon have been used to reduce the number of independent Lorentz operators:
\begin{equation}
    \gamma^\alpha u_\alpha(p_1) = 0\ ,\quad  p_1^\alpha u_\alpha(p_1)=0, \quad 
    \slashed{p}_1 u_\alpha(p_1)=m_\Delta u_\alpha(p_1), \quad 
    \bar u_N(p_2) \slashed{p}_2 = m_N \bar u_N(p_2)\ . 
    \label{eq:EOMs}
\end{equation}

In the case of $q_1 = q_2 \equiv q$ (i.e., $t=u$), the symmetric hadronic amplitude $T_S^\alpha$ in Eq.~\eqref{eq:def-Amp-S} can be further simplified to 
\begin{equation}
      T_S^\alpha(q) =
      T q^\alpha + T^\prime q^\alpha\gamma_5
        \label{eq:def-Amp-TTp}
\end{equation}
by applying the Gordon identity, where 
\begin{subequations}
    \begin{eqnarray}
        T &=& T_1 + T_2 + \frac{m_\Delta - m_N}{2}(T_3 + T_4) = 2 T_1 + (m_\Delta - m_N)T_3~,\\
         T^\prime &=& T_1^\prime + T_2^\prime - \frac{m_\Delta + m_N}{2}(T_3^\prime + T_4^\prime) = 2 T_1^\prime - (m_\Delta + m_N)T_3^\prime~.
    \end{eqnarray}
\label{eq:T-Tprime}
\end{subequations}
\!\!We refer to the quantities $T$ and $T^\prime$ as neutrinoless transition form factors (TFFs), as they parameterize the hadronic structure of the neutrinoless $\Delta\to p$ transition in the kinematic configuration with collinear electrons. 
They provide a convenient, model-independent description of the underlying QCD dynamics at low energies.

\subsection{Unrenormalized one-loop amplitude and power counting}
\label{subsec:PCB}
\begin{figure}[t]
\includegraphics*[width=0.225\linewidth]{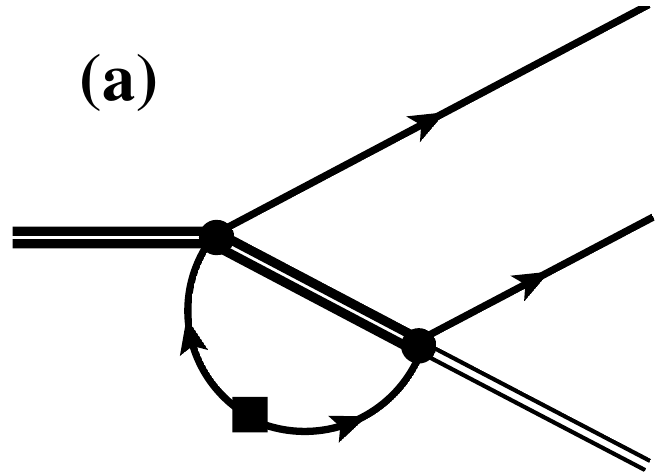} \quad
\includegraphics*[width=0.225\linewidth]{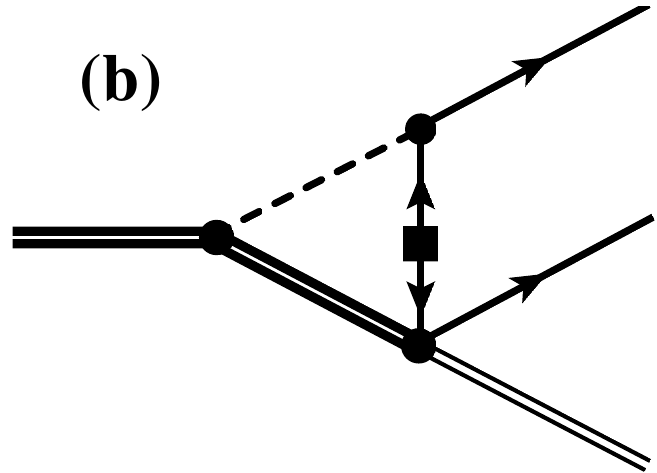} \quad
\includegraphics*[width=0.225\linewidth]{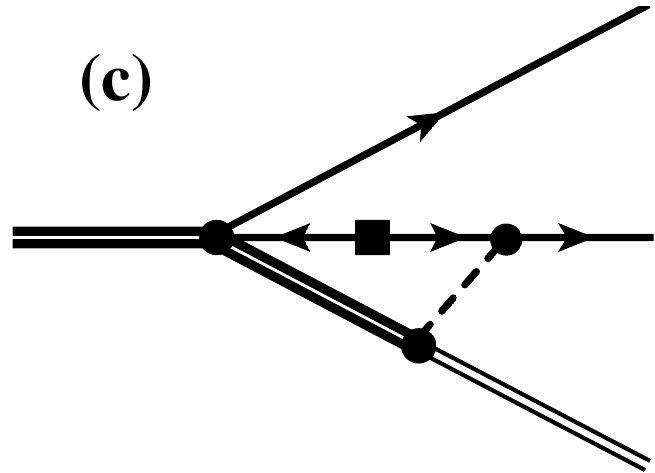} \quad
\includegraphics*[width=0.225\linewidth]{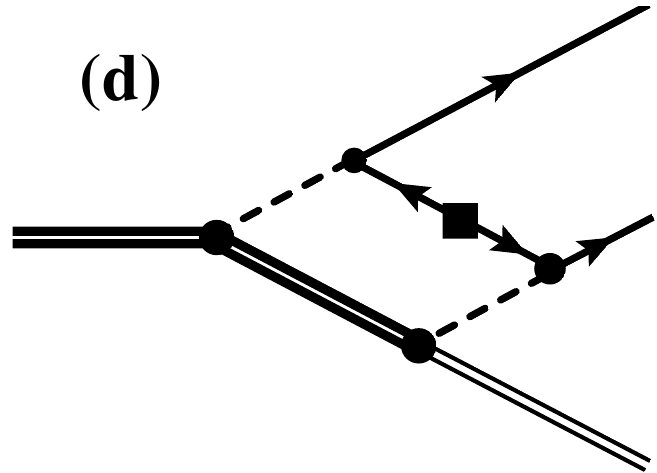}
\includegraphics*[width=0.225\linewidth]{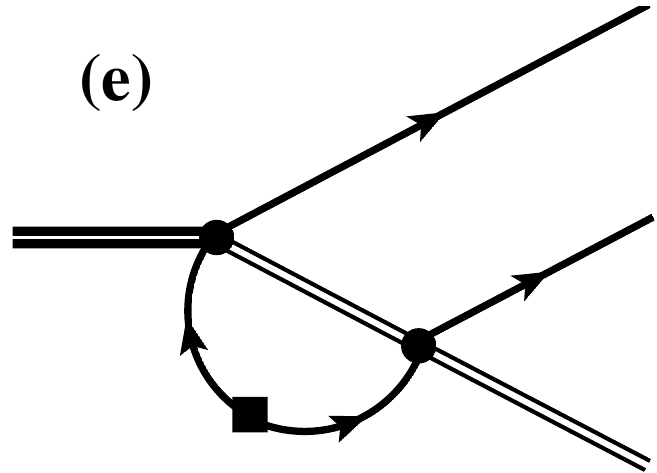} \quad
\includegraphics*[width=0.225\linewidth]{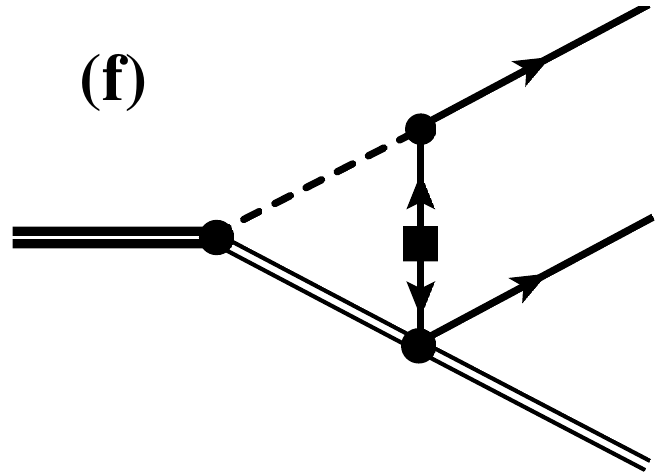} \quad
\includegraphics*[width=0.225\linewidth]{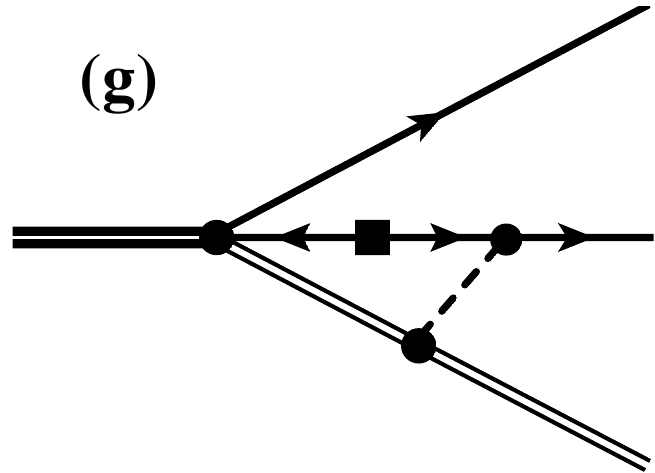} \quad 
\includegraphics*[width=0.225\linewidth]{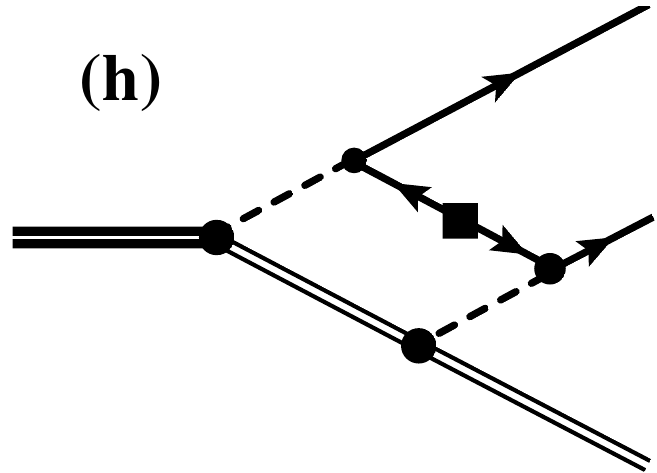}
\caption{
One-loop Feynman diagrams contributing to the $0\nu\beta\beta$ decay $\Delta^- \to p e^-e^-$.
Leptons, pion, nucleons and deltas are represented by solid, dashed, thin double and thick double lines, in order. The $\Delta L=2$ vertex, proportional to the effective Majorana neutrino mass, is indicated by a black solid square. Crossed diagrams, obtained by interchanging the two final-state electrons, are not shown explicitly.
}
\label{fig:Feyn-LR}
\end{figure}
In $\chi$EFT, the LO contribution to the decay $\Delta^- \to p e^-e^-$ arises at one loop. The one-loop Feynman diagrams relevant to our calculation are displayed in Fig.~\ref{fig:Feyn-LR}. The corresponding crossed diagrams, obtainable by interchanging the final electrons, are not shown explicitly. According to the standard Weinberg power counting rule~\cite{Weinberg:1991um}, an interaction vertex with chiral order $i$ scales as $\mathcal{O}(p^i)$, a pion propagator is counted as $\mathcal{O}(p^{-2})$ and a nucleon or delta propagator as $\mathcal{O}(p^{-1})$.
Hence, all diagrams in Fig.~\ref{fig:Feyn-LR} are of $\mathcal{O}(p^3)$. Furthermore, a charged lepton in an interaction vertex is assigned to be $\mathcal{O}(p)$, following Ref.~\cite{Cirigliano:2017tvr}. Therefore, the symmetric hadronic amplitude $T_S^\alpha$, defined in Eq.~\eqref{eq:def-Amp}, receives contributions from $\mathcal{O}(p)$.

In the process $\Delta^-\to pe^-e^-$, the baryon masses $m_N$ and $m_\Delta$ are hard scales. However, they have a small mass difference $\delta = m_\Delta - m_N$, which can be assigned as $\mathcal{O}(p)$ in the SSE scheme. We take the four-momenta $q_1,q_2$ of the outgoing electrons and the three-momenta ${\bm p}_1,{\bm p}_2$ of the nucleons as small quantities, scaling as $\mathcal{O}(p)$. Since the Mandelstam variables $t$ and $u$ lie within the range $m_N^2 \leqslant t,u \leqslant m_\Delta^2$, the combinations $t - m_N^2$ and $u - m_N^2$ can be counted as $\mathcal{O}(p)$, as the counting for the inverse of the nucleon propagator.

In Eq.~\eqref{eq:def-Amp-S}, the eight Lorentz coefficients $T_i^{(\prime)} (i = 1, 2, 3, 4)$ have different chiral orders. 
Given that the decay amplitude is $\mathcal{O}(p^3)$, the power counting of $T_i^{(\prime)}$ is summarized in Table~\ref{tab:Ti}.  
Specifically, since $q_1\sim q_2\sim \mathcal{O}(p)$, $T_{1,2}$ behave as $\mathcal{O}(p^0)$, and $T_{3,4}^{(\prime)}$  as $\mathcal{O}(p^{-1})$. The powers of $T_{1,2}^\prime$ are one order lower than those of $T_{1,2}$, owing to the presence of $\gamma_5$ sandwiched between the baryon spinors $\bar u_N(p_2)$ and $u_\alpha(p_1)$. 
\begin{table}[tb]
    \centering
    \caption{Power counting of the eight Lorentz coefficients and the two TFFs in our one-loop calculation of the decay amplitude. 
    }
    \begin{tabular}{c|cc|cc|cc|cc|c|c}
        &$~T_1~$ &$~T_2~$ &$~T_3~$ &$~T_4~$ &$~T_1^\prime~$ &$~T_2^\prime~$ &$~T_3^\prime~$ &$~T_4^\prime~$  & $T$ & $T^\prime$\\  \hline
        power counting
        & \multicolumn{2}{c|}{$p^0$}   &\multicolumn{2}{c|}{$p^{-1}$} & \multicolumn{2}{c|}{$p^{-1}$}   &\multicolumn{2}{c|}{$p^{-1}$} &$~~p^{0}~~$  &$~~p^{-1}~~$ 
    \end{tabular}
    \label{tab:Ti}
\end{table}

We employ \texttt{FeynCalc}~\cite{Mertig:1990an, Shtabovenko:2016sxi, Shtabovenko:2020gxv} and \texttt{FeynHelpers}~\cite{Shtabovenko:2016whf} to carry out  symbolic computation of the one-loop Feynman diagrams in Fig.~\ref{fig:Feyn-LR}. The resultant loop amplitudes can be further expressed as linear combinations of one-loop scalar integrals by implementing the Passarino–Veltman reduction procedure~\cite{Passarino:1978jh}, which are subsequently evaluated in dimensional regularization using \texttt{Package-X}~\cite{Patel:2016fam}. 
This computational framework enables an efficient and fully analytic treatment of the one-loop amplitudes. In this manner, we obtain the analytical expressions for the Lorentz coefficients $T_{i}^{(\prime)}$ ($i=1,2,3,4$). The full expressions are, however, too lengthy to be presented here.


\subsection{Counterterms}
\label{subsec:Amp-CC}

\begin{figure}[htb]
\includegraphics*[width=0.25\linewidth]{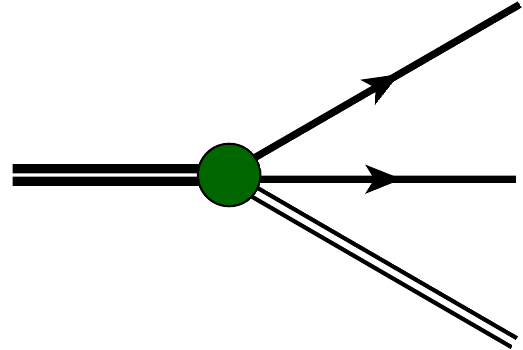} 
\caption{
Feynman diagram for the counterterm of $\Delta^- \to p \,e^-e^-$.}
\label{fig:Feyn-CC}
\end{figure}

In this subsection, the short-range counterterms are constructed using a bottom-up approach. The corresponding counterterm diagram is presented in Fig.~\ref{fig:Feyn-CC}.
The counterterms are needed to cancel the ultraviolet (UV) divergences of the one-loop symmetric amplitude $T_S^\alpha$ obtained in the previous subsection. To that end, the counterterms must contain the leptonic structure $\bar e_L C \bar e_L^{T}$, the same one that multiplies $T_S^\alpha$ in Eq.~\eqref{eq:def-Amp}. Note that the same leptonic structure was also adopted in the counterterms for the $nn \to ppe^-e^-$ process discussed in Ref.~\cite{Cirigliano:2017tvr}.

Since any counterterm operator without derivatives acting on the fermion fields must contains a factor of $\gamma^\mu \Psi_\mu^i$, which vanishes upon employing the EOM for the $\Delta$, nonvanishing counterterms must involve partial derivatives acting on the matter fields. In Appendix~\ref{sec.counterterms}, we present a systematic construction of these operators with one, two, and three derivatives. Redundant structures are removed using the EOMs. The resulting set of counterterms is sufficient to cancel the UV divergences arising from the one-loop symmetric hadronic amplitude $T_S^\alpha$ considered in this work.

Specifically, the required counterterm Lagrangian with correct transformation properties can be written as
\begin{subequations}
\begin{eqnarray}
 \mathcal{L}_{\mathrm{ct}} &=& \mathcal{L}_{1,\mathrm{ct}} +\mathcal{L}_{\mathrm{2,ct}} +\mathcal{L}_{3,\mathrm{ct}} \ ,\\
\mathcal{L}_{1,\mathrm{ct}} &=&
            G\bar N(\overleftarrow{\partial}^\mu - \Gamma^\mu) (h_1+h_1^\prime\gamma_5) u^\dagger \tau^+u\tau^iu^\dagger\tau^+u\Psi_\mu^i \bar e_LC \bar e_L^T + \text{H.c.}\ ,\label{eq.CT1} \\
\mathcal{L}_{2,\mathrm{ct}} &=&
             G\bar N \gamma_\nu(h_2+h_2^\prime\gamma_5)u^\dagger \tau^+u\tau^iu^\dagger\tau^+u\Psi_\mu^i  (\partial^\mu\bar e_L) C (\partial^\nu\bar e_L^T)  + \text{H.c.}\ ,\label{eq.CT2}\\
\mathcal{L}_{3,\mathrm{ct}}
             &=&
             G\bar N(\overleftarrow{\partial}_\sigma - \Gamma_\sigma)\big[ g^{\sigma\mu}g_{\nu\rho}(h_{3a}+h_{3a}^\prime\gamma_5)  
             + g_\nu^\sigma\gamma^\mu\gamma_\rho (h_{3b}+h_{3b}^\prime\gamma_5)  \big] \notag\\
             &&  \hspace{0.5cm}
             \times u^\dagger \tau^+u\tau^iu^\dagger\tau^+u \Psi_\mu^i (\partial^\nu\bar e_L) C (\partial^\rho\bar e_L^T) + \text{H.c.}\ , \label{eq.CT3}
\end{eqnarray}
\end{subequations}
where the subscript $i$ on each $\mathcal{L}_{i,\mathrm{ct}}$ denotes the number of partial derivatives acting on the fermion fields. The unknown LECs $h_1^{(\prime)}$, $h_2^{(\prime)}$ and $h_{3a,3b}^{(\prime)}$ have mass dimensions of $0$, $-1$ and $-2$,
respectively. 

With the Lagrangians in Eqs.~\eqref{eq.CT1}, \eqref{eq.CT2}, and \eqref{eq.CT3}, one obtains the counterterm amplitude for $\Delta^- \to p e^-e^-$, which reads
\begin{align}
       \mathcal{M}_\mathrm{ct} = 
        \sqrt{2}i\, G\bar{u}_N\! \left(p_2\right)  
        \big[&T_{1,\mathrm{ct}}\, q_1^\alpha + T_{2,\mathrm{ct}}\, q_2^\alpha + T_{3,\mathrm{ct}}\, q_1^\alpha\slashed{q}_2 + T_{4,\mathrm{ct}}\, q_2^\alpha\slashed{q}_1 + T_{1,\mathrm{ct}}^\prime\, q_1^\alpha\gamma_5 \notag\\
        & +
        T_{2,\mathrm{ct}}^\prime\, q_2^\alpha\gamma_5 + T_{3,\mathrm{ct}}^\prime\, q_1^\alpha\slashed{q}_1\gamma_5 + T_{4,\mathrm{ct}}^\prime\, q_2^\alpha\slashed{q}_1\gamma_5 \big] u_\alpha\!\left(p_1\right)
        \bar{u}_L(q_1) C\bar{u}_L^T(q_2) \ , 
\label{eq:def-Amp-CC}
\end{align}
with the coefficients given by
\begin{subequations}
\begin{eqnarray}
        T_{1,\mathrm{ct}}^{(\prime)} 
        &=&
        2 h_1^{(\prime)}  -  (h_{3a}^{(\prime)}+ h_{3b}^{(\prime)}) m_N^2 - h_{3a}^{(\prime)}  m_\Delta^2 +  (h_{3a}^{(\prime)}  + h_{3b}^{(\prime)} ) t + h_{3a}^{(\prime)}  u\ , \\
        T_{2,\mathrm{ct}}^{(\prime)} &=& T_{1,\mathrm{ct}}^{(\prime)} ( t\leftrightarrow u)\ ,  \label{eq:T2cc}\\
        T_{3,\mathrm{ct}}^{(\prime)}  &=&  T_{4,\mathrm{ct}}^{(\prime)} =
        -i h_2^{(\prime)}\ . 
\end{eqnarray}
\end{subequations}
Eq.~\eqref{eq:T2cc} indicates that $T_{2,\mathrm{ct}}^{(\prime)}$ is obtained from $T_{1,\mathrm{ct}}^{(\prime)} $ by interchanging $t$ and $u$. 
According to the naive power counting rule, the counterterm in Eq.~\eqref{eq:def-Amp-CC} is $\mathcal{O}(p^3)$, where 
the coefficients $T_{i,\mathrm{ct}}^{(\prime)}$ contribute at the same order as those listed in Table~\ref{tab:Ti}. 
They can be written in the following chiral-expansion form, 
\begin{subequations}
    \begin{eqnarray}
        T_{1,\mathrm{ct}}^{(\prime)} 
        &=&
        T_{2,\mathrm{ct}}^{(\prime)} 
        =
        2 h_1^{(\prime)}  + \mathcal{O}(p^1)\ , \\
        T_{3,\mathrm{ct}}^{(\prime)}  &=&  T_{4,\mathrm{ct}}^{(\prime)} =
        0 + \mathcal{O}(p^0)\ .
    \end{eqnarray}
\end{subequations}
In the case of $t=u$, one readily obtains the counterterms for the TFFs $T$ and $T^\prime$ defined in Eqs.~\eqref{eq:T-Tprime},
\begin{subequations}
    \begin{eqnarray}
        T_\mathrm{ct}
        &=&
        2\sqrt{2} i[2 h_1  - (h_{3a}+ h_{3b}) m_N^2 - h_{3a}  m_\Delta^2 +  (2h_{3a}  + h_{3b} ) t] -i h_2 (m_\Delta - m_N) \nonumber\\
        &=& 
        4\sqrt{2}i h_1 + \mathcal{O}(p^1)\ , \\
        T_{\mathrm{ct}}^{\prime}  
        &=& 
         2\sqrt{2} i[2 h_1^{\prime}  -  (h_{3a}^{\prime}+ h_{3b}^{\prime}) m_N^2 - h_{3a}^{\prime}  m_\Delta^2 +  (2h_{3a}^{\prime}  + h_{3b}^{\prime} )t ] + i h_2^{\prime}  (m_\Delta + m_N) \nonumber\\
         &=&
         0+ \mathcal{O}(p^0)\ .
    \end{eqnarray}
\end{subequations}

\subsection{Ultraviolet divergences and renormalization}
\label{subsec:Amp-UV}
Having established the necessary counterterms, we now proceed to address the cancellation of UV divergences. The UV-divergent terms of the symmetric hadronic amplitude $T_S^\alpha$ can be expressed as 
\begin{align}
T_{S}^{\alpha,\mathrm{UV}} = -R
        \big[&T_1^{(\text{UV})} q_1^\alpha + T_2^{(\text{UV})}  q_2^\mu + T_3^{(\text{UV})} q_1^\alpha\slashed{q}_2 + T_4^{(\text{UV})} q_2^\alpha\slashed{q}_1 \nonumber\\
         &+
        T_1^{\prime{(\text{UV})}} q_1^\alpha\gamma_5 + T_2^{\prime{(\text{UV})}} q_2^\alpha\gamma_5 + T_3^{\prime{(\text{UV})}} q_1^\alpha\slashed{q}_2\gamma_5 + T_4^{\prime{(\text{UV})}} q_2^\alpha\slashed{q}_1\gamma_5\big]\ ,
\label{eq:divergences}
\end{align}
where $R = - 2/(4-d)) + \gamma_E -1 - \log(4\pi)$, with $d$ the dimension of spacetime and $\gamma_E$ the Euler constant. The coefficients from the one-loop diagrams of Fig.~\ref{fig:Feyn-LR} are given by  
\begin{subequations}
    \begin{eqnarray}
        T_{1}^{(\text{UV})} &=& \frac{25m_\Delta^2 -6 m_N^2 -3m_\Delta m_N -t +3u +12 M_\pi^2}{576\sqrt{2} \, \pi^2  m_\Delta^2  } \ ,  \\
        T_{3}^{(\text{UV})} &=&  T_{4}^{(\text{UV})} 
        =\frac{m_N-m_\Delta}{144\sqrt{2} \, \pi^2  m_\Delta^2  } \ ,
        \\
        T_{1}^{\prime(\text{UV})} &=& 
        \frac{108 g_A\, m_\Delta^2 + g_1 (17 m_\Delta^2-5 m_N^2 +14m_\Delta m_N +7t +5u +6 M_\pi^2)}{1728\sqrt{2} \, \pi^2  m_\Delta^2  } \ ,\\
        T_{3}^{\prime(\text{UV})} &=&  T_{4}^{\prime(\text{UV})} 
        =\frac{g_1(3m_\Delta-2m_N)}{864\sqrt{2} \, \pi^2  m_\Delta^2  }\ . 
    \end{eqnarray}
    \label{eq:UV}
\end{subequations}
The UV divergent terms $T_{2}^{(\prime)(\text{UV})}$ can be obtained from $T_{1}^{(\prime)(\text{UV})} $ by interchanging $t$ and $u$. 
These UV divergences can be canceled by the $\Delta^-\to p e^-e^-$ counterterm given in Eq.~\eqref{eq:def-Amp-CC}. To that end, the LECs are split as
\begin{equation}
         h_i^{(\prime)} = \delta h_i^{(\prime)} R +  h_{i}^{(\prime)R} , \quad i =1, 2, 3a, 3b\ ,  
         \label{eq:hi}
\end{equation}
where the first term $\delta h_i^{(\prime)} R$ exactly cancels the UV divergences from the loop diagrams. 
At the orders listed in Table~\ref{tab:Ti}, 
\begin{equation}
        T_{1}^{(\text{UV})} = T_{2}^{(\text{UV})}  = 
        \frac{1}{32\sqrt{2}\,\pi^2}\ , 
    \label{eq:UVLO}
\end{equation}
and all other coefficients are 0. 
The shifts $\delta h_i^{(\prime)}$ are determined to be
\begin{align}
        \delta h_1 =
               \frac{-i}{128 \pi^2} , \quad
        \delta h_{i}^{(\prime)} = 
                0  \quad \text{for } i = 2,3a,3b\ .
    \label{eq:CC-deltah}
\end{align}
The remaining parts $h_i^{(\prime)R}$ in Eq.~\eqref{eq:hi} correspond to the UV-renormalized LECs, which may be determined when future lattice QCD calculations are available.


It is well known that the notable power-counting-breaking (PCB) problem arises from the presence of internal matter-field propagators in loops~\cite{Gasser:1987rb}. For $\Delta^-\to pe^-e^-$, however, 
no PCB terms appear once the Lorentz decomposition of $T_S^\alpha$ is organized as in Eq.~\eqref{eq:def-Amp-S}. Other decompositions may still generate intermediate PCB pieces, but the physical amplitude can be cast into a PCB-free form.
Consequently, a finite renormalization procedure, using, e.g., the extended-on-mass-shell scheme~\cite{Fuchs:2003qc}, is unnecessary here, in contrast to the case of hyperon $0\nu\beta\beta$ decays~\cite{Zhao:2026klv}.

\section{Pion mass dependence}
\label{sec:pionmass}
\begin{figure}[t]
\includegraphics*[width=0.6\linewidth]{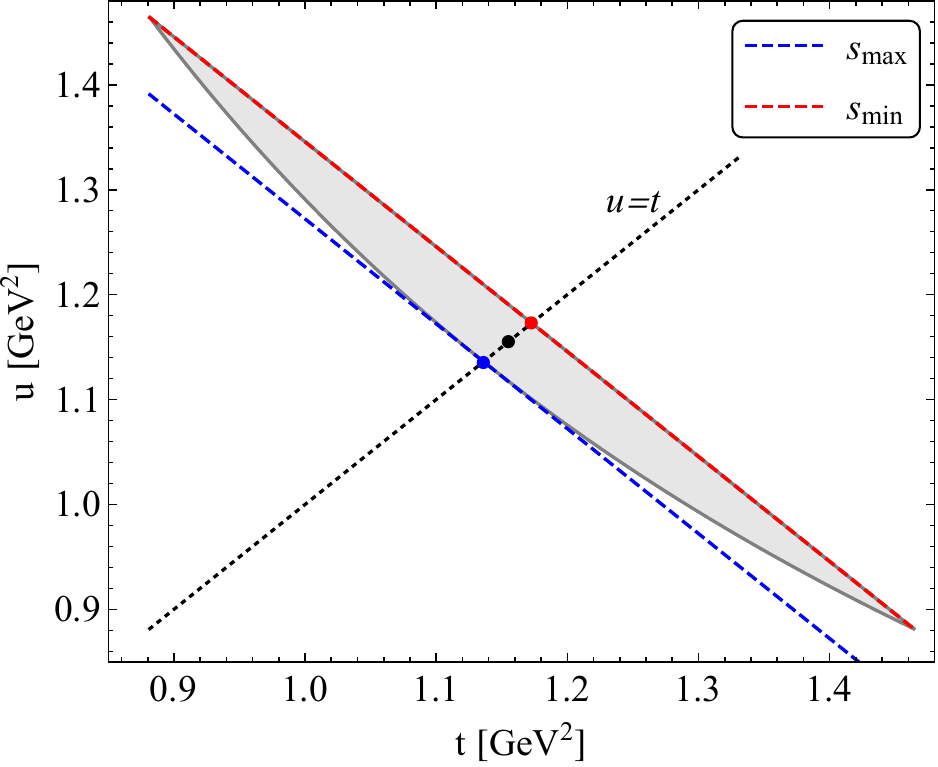} 
\caption{Kinematically allowed region of the $\Delta^- \to p e^-e^-$ phase space in the $u-t$ plane. The red dashed line corresponding to $s_\mathrm{min}$ coincides with the upper bound of $u$ in terms of $t$, while the blue dashed line corresponding to $s_\mathrm{max}$ intersects the allowed region only at the blue point with $u=t$. The black dashed line corresponds to the case $u = t$. The black and red points denote the kinematic configuration with $u = t$ at $s = s_{\max}/2$ and $s = s_{\min}$, respectively. }
\label{fig:Dalitz}
\end{figure}

In this section, we present the pion mass dependence of the symmetric amplitude $T_S^\alpha$ defined in Eq.~\eqref{eq:def-Amp}. 
In our numerical computation, the electron mass is set to zero, since its value is much smaller than the pion mass. The nucleon mass is taken to be the average of the neutron and proton masses, while the $\Delta$ mass is set to be $m_\Delta=1.210$~GeV~\cite{ParticleDataGroup:2024cfk}. For $\Delta^-\to pe^-e^-$, the allowed kinematical region in the $t-u$ plane is shown as the gray area in Fig.~\ref{fig:Dalitz}. 
The black short-dashed line corresponds to $t=u$. To illustrate the pion mass dependence, we select three representative kinematical points along this line:
\begin{itemize}
    \item Point A: 
    $s=s_{\mathrm{max}} = \delta^2$, for which the three-body phase-space constraint forces $t$ and $u$ to be equal to $m_\Delta m_N$. Namely, for this preferred $s$ value, the allowed kinematic region collapses to the single blue point in Fig.~\ref{fig:Dalitz}.
    \item Point B:
    $s=s_{\mathrm{max}}/2$, $t= u = (m_\Delta + m_N)^2/4$, shown as the black point in Fig.~\ref{fig:Dalitz}.
    \item Point C:
    $s=s_{\mathrm{min}} = 0$, $t= u = (m_\Delta^2 + m_N^2)/2$, shown as the red point in Fig.~\ref{fig:Dalitz}.
\end{itemize}
We can ignore the pion mass dependence of $m_\Delta$ and $m_N$ as this amounts to a higher order correction. Consequently, 
the mass difference $\delta$ is a constant independent of $M_\pi$.

\subsection{Chiral extrapolation of decay amplitude}
\label{subsec:pionmass}
In the kinematic configuration $t=u$, the antisymmetric hadronic amplitude $T_{A}^{\alpha,\mu\nu}$ in Eq.~\eqref{eq:def-Amp} vanishes and only the symmetric part $T_S^\alpha$ survives. $T_S^\alpha$ can be generically decomposed as Eq.~\eqref{eq:def-Amp-S}. With $t=u$, it can be further simplified to the form in Eq.~\eqref{eq:def-Amp-TTp}, which involves only two independent Lorentz structures, i.e., the TFFs $T$ and $T^\prime$.
$T$ involves no LECs and originates only from diagrams (a), (c), (e) and (f) of Fig~\ref{fig:Feyn-LR}, whereas $T^\prime$ depends on the LECs $g_A$ and $g_1$ and receives contributions from all the eight Feynman diagrams. Note that no pion propagators occur in diagrams (a) and (e). Therefore, their contributions are pion-mass independent when the pion-mass dependence of the nucleon and delta masses are neglected. Furthermore, they are weakly dependent on $s$, since 
\begin{subequations}
    \begin{eqnarray}
        T_{(a)} &\simeq& -0.003\ , \hspace{1.62cm}  T_{(a)} ^\prime \simeq -0.005\ , \label{eq.tffs.a} \\
        T_{(e)} &\simeq& 0.010 +0.006 i\ , \quad 
        T_{(e)} ^\prime \simeq 0.013 +0.0075i\ ,    \label{eq.tffs.e} 
        \end{eqnarray}
\end{subequations}
for any $s\in [s_\mathrm{min}, s_\mathrm{max}]$.

\begin{figure}[t]
\includegraphics*[width=0.49\linewidth]{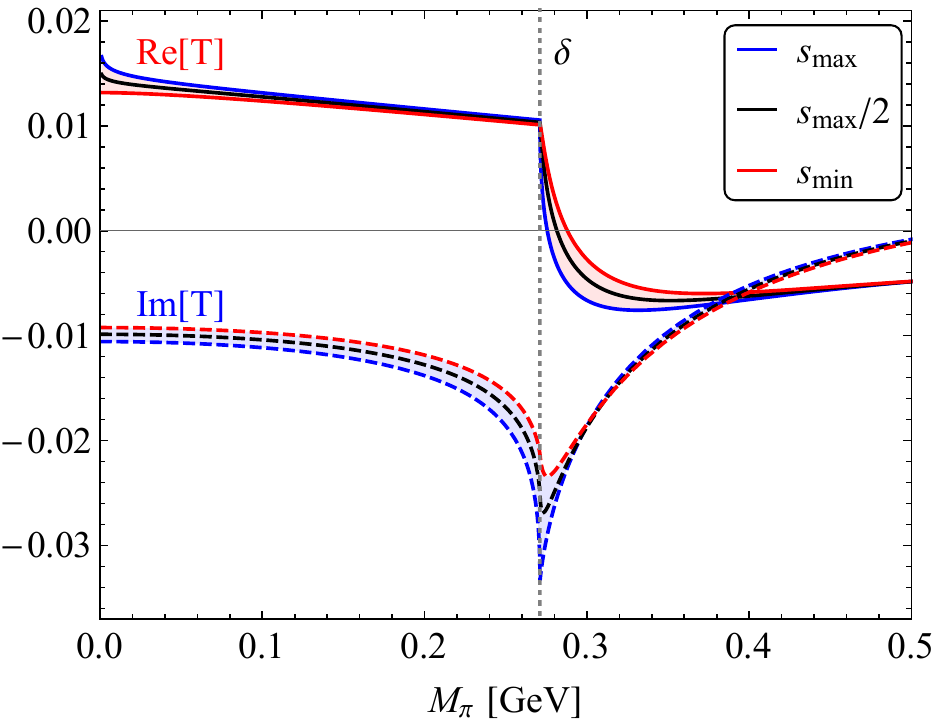} 
\includegraphics*[width=0.48\linewidth]{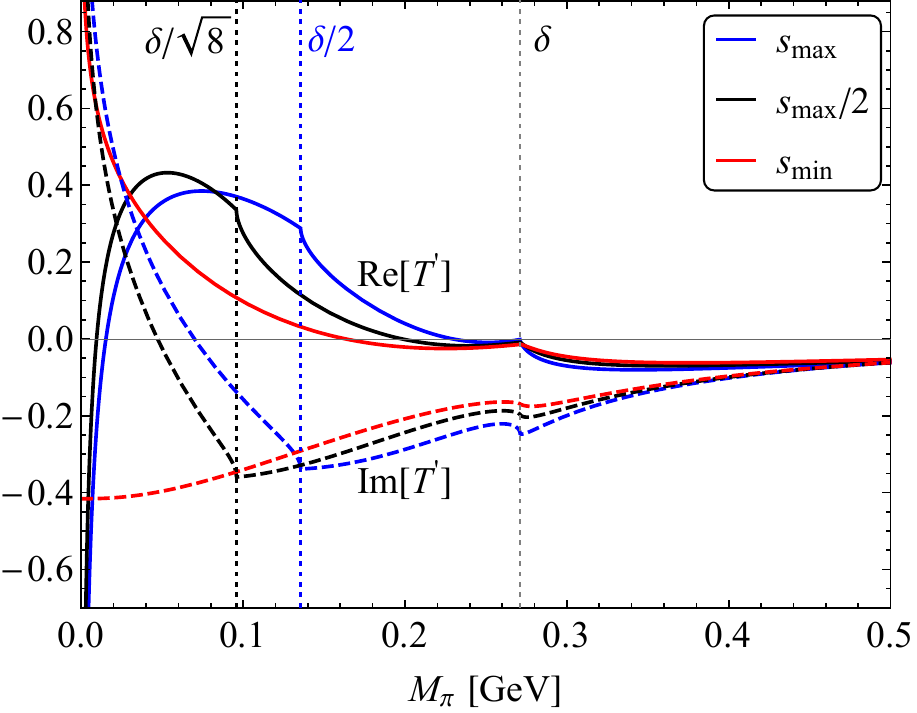} 
\caption{
Pion mass dependence of $T$ (left) and $T^\prime$ (right) defined in Eqs.~\eqref{eq:T-Tprime}. 
The solid and dashed lines are the real and imaginary parts for $s = s_\mathrm{max}$ (blue lines), $s = s_\mathrm{max}/2$ (black lines), and $s = s_\mathrm{min} =0$ (red lines) in the case of $u=t$. 
The vertical gray dotted line is the mass difference $\delta = m_\Delta - m_N$.
The vertical blue and black dotted lines are the positions where cusps arise in diagrams (d) and (h) shown in Fig.~\ref{fig:Feyn-LR}.
The light-red (real part) and light-blue (imagianry part) bands indicate the regions between the curves corresponding to $s_{\mathrm{max}}$ and $s_{\mathrm{min}}$ on the left panel. 
}
\label{fig:T-Tprime}
\end{figure}
%
The pion mass dependence of $T$ and $T^\prime$ is presented in Fig.~\ref{fig:T-Tprime} for the three aforementioned kinematical points. The TFFs from diagrams (a), (b) and (c) are real. All other diagrams yield complex TFFs. 
The imaginary part arises when at least two internal propagators can be put on-shell simultaneously. The enhancements of imaginary parts, due to threshold cusps and triangle singularities (for a review, we refer to Ref.~\cite{Guo:2019twa}), will be examined in detail in the next subsection.

\begin{figure}[t]
\includegraphics*[width=0.5\linewidth]{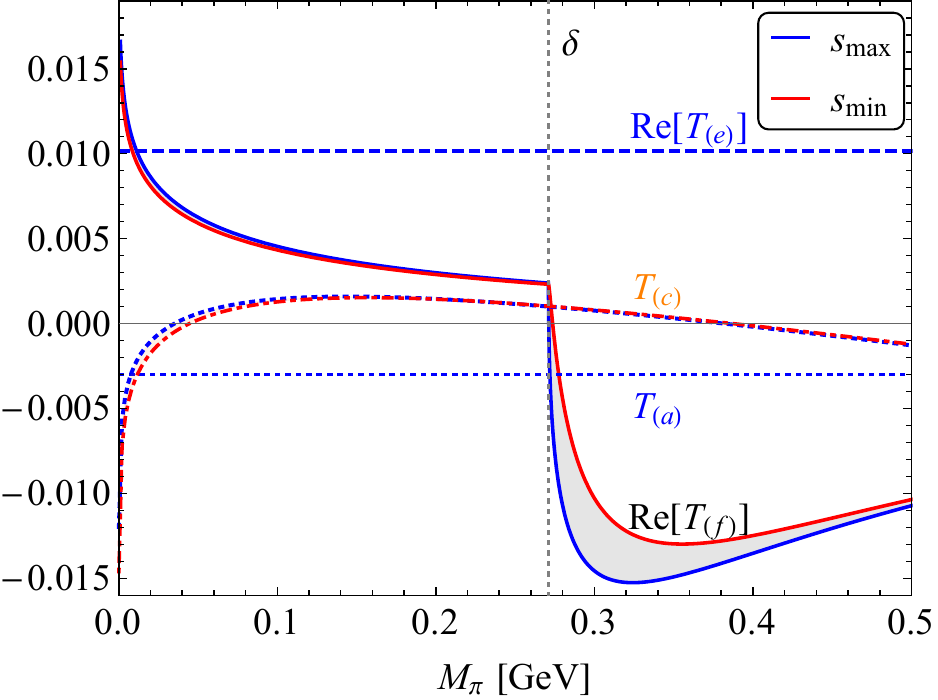} 
\caption{
Pion mass dependence of the real part of $T$ from diagrams (a) (blue dotted line), (c) (orange band), (e) (blue dashed line) and (f) (gray band). 
The blue and red lines are for $s = s_\mathrm{max}$ and $s = s_\mathrm{min}$, respectively.}
\label{fig:pion-mass-ReT}
\end{figure}

As can be seen from the left panel of Fig.~\ref{fig:T-Tprime}, the pion-mass dependence of $T$ for different values of $s$ remains comparable in magnitude and shows weak sensitivity to the specific choice of $s$, except near the threshold at $M_\pi = \delta$. On the one hand, the imaginary part Im[$T$] receives contributions only from diagrams (e) and (f), where diagram (e) provides a constant contribution. 
The pion mass dependence of Im[$T$] only arises from diagram (f).
The magnitude is greatly enhanced near $M_\pi=\delta$ due to the presence of a  triangle singularity, which is the leading Landau singularity~\cite{Landau:1959fi} of triangle diagram. On the other hand, the real part Re[$T$] gets contributions from diagrams (a), (c), (e) and (f), as shown individually in Fig.~\ref{fig:pion-mass-ReT}.
Likewise, there exists a large enhancement for $M_\pi >\delta$, caused by a nearby triangle singularity stemming from diagram (f). 
In the chiral limit ($M_\pi \to 0$), diagrams (c) and (f) individually yield logarithmic divergences $\propto \log M_\pi$ with opposite signs. Their near cancelation leaves a residual $\log M_\pi$ divergence whose coefficient is highly suppressed. Consequently, this divergence is not visible in the left panel of Fig.~\ref{fig:T-Tprime}, particularly for the red curve ($s = s_{\mathrm{min}}$).

\begin{figure}[t]
\includegraphics*[width=0.485\linewidth]{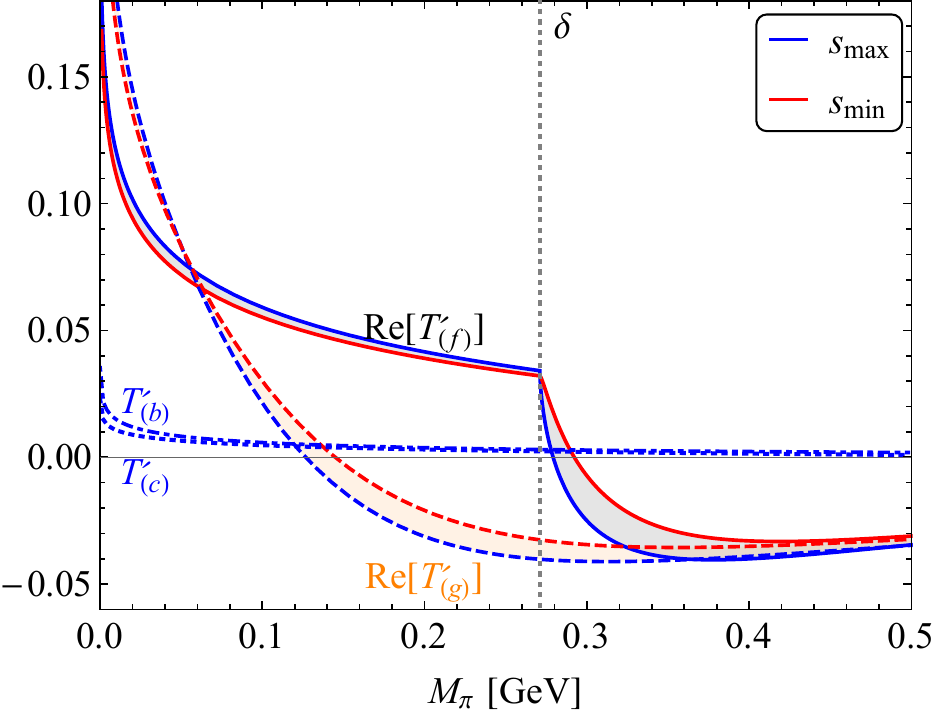} 
\includegraphics*[width=0.482\linewidth]{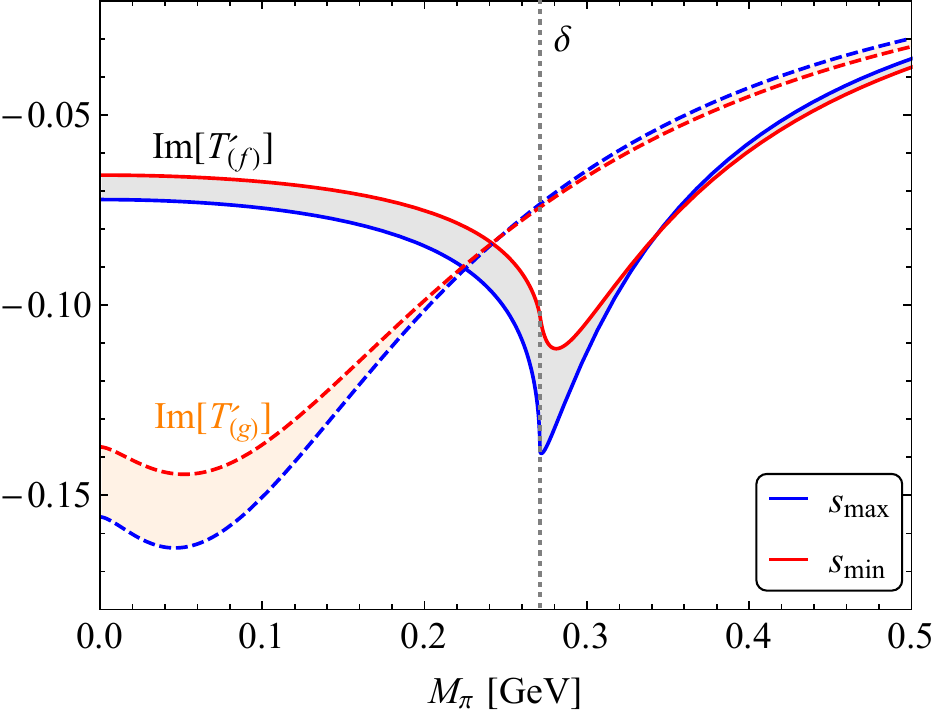} 
\caption{
Pion mass dependence of the real (left) and imaginary (right) parts of $T^\prime$ from diagrams (b) (dot-dashed line), (c) (dotted line), (f) (gray band) and (g) (orange band). 
The blue and red lines are for $s = s_\mathrm{max}$ and $s = s_\mathrm{min} =0$. The imaginary parts from diagrams (b) and (c) are 0.}
\label{fig:pion-mass-Tp-fg}
\end{figure}
\begin{figure}[t]
\includegraphics*[width=0.49\linewidth]{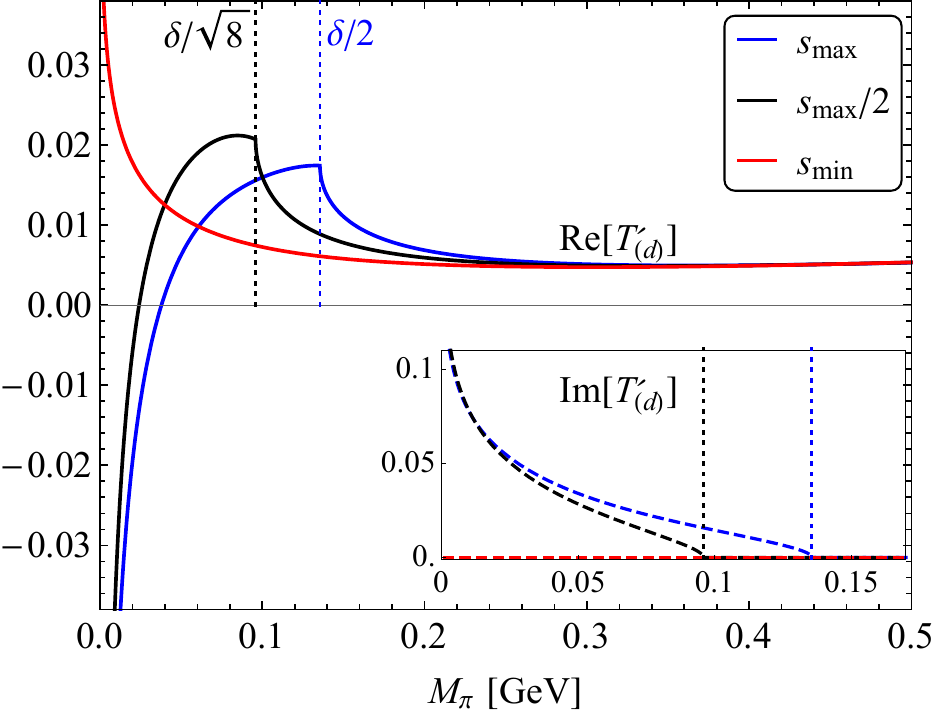} 
\includegraphics*[width=0.48\linewidth]{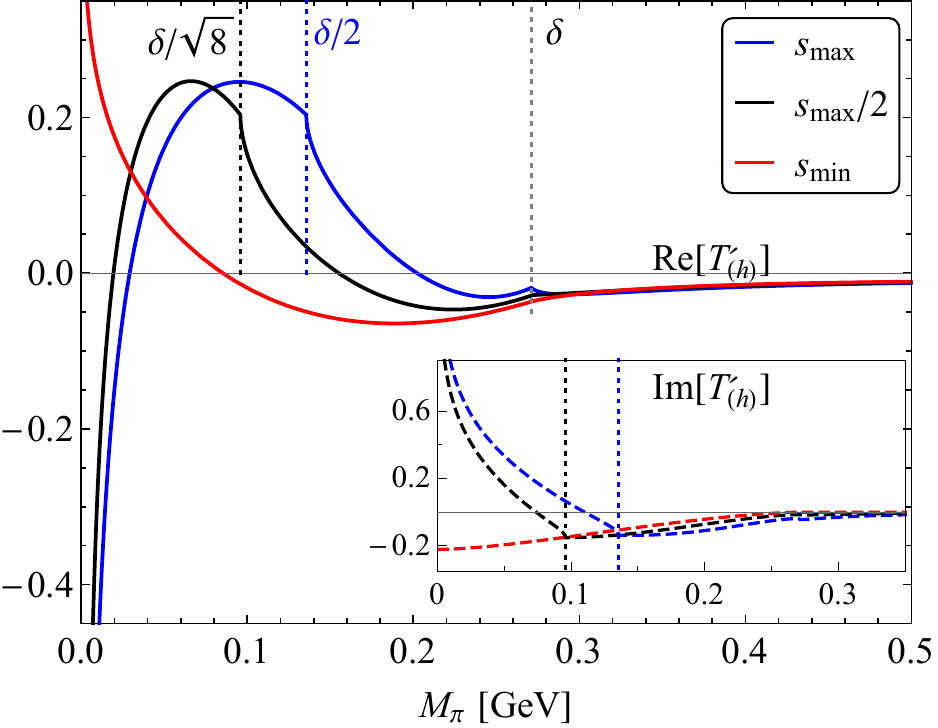} 
\caption{
Pion mass dependence of $T^{\prime}$ with $t=u$ from diagrams (d) (left) and (h) (right). The imaginary parts are shown as the insets.}
\label{fig:pion-mass-Tp-dh}
\end{figure}
\begin{figure}[t]
\includegraphics*[width=0.49\linewidth]{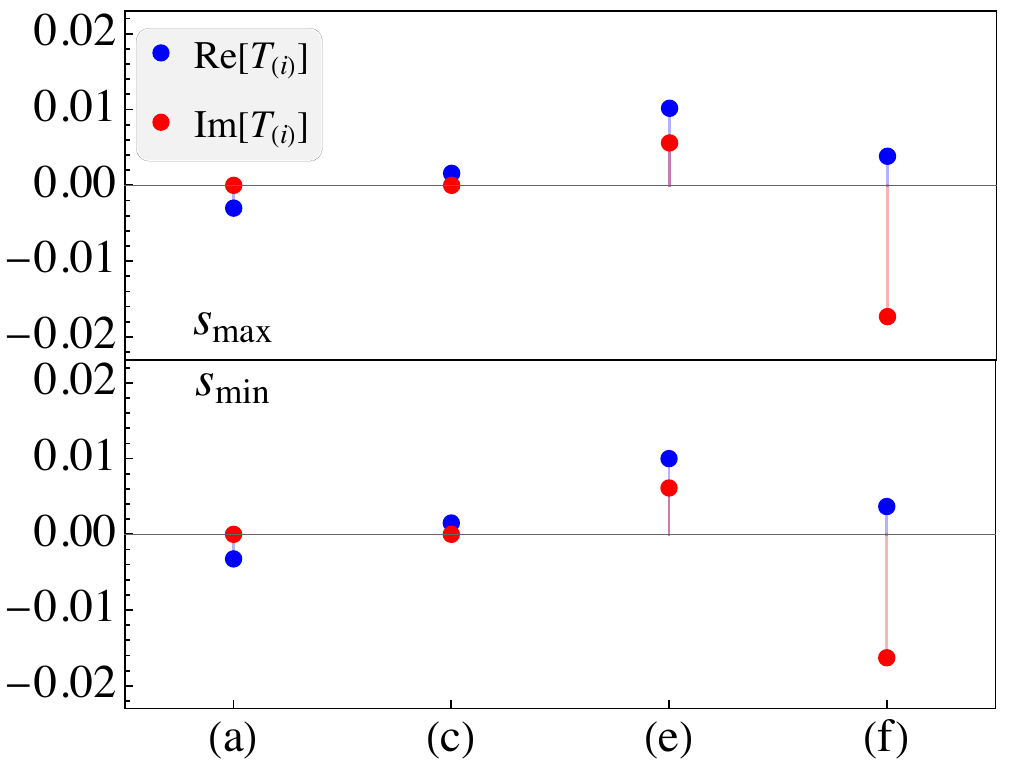} 
\includegraphics*[width=0.48\linewidth]{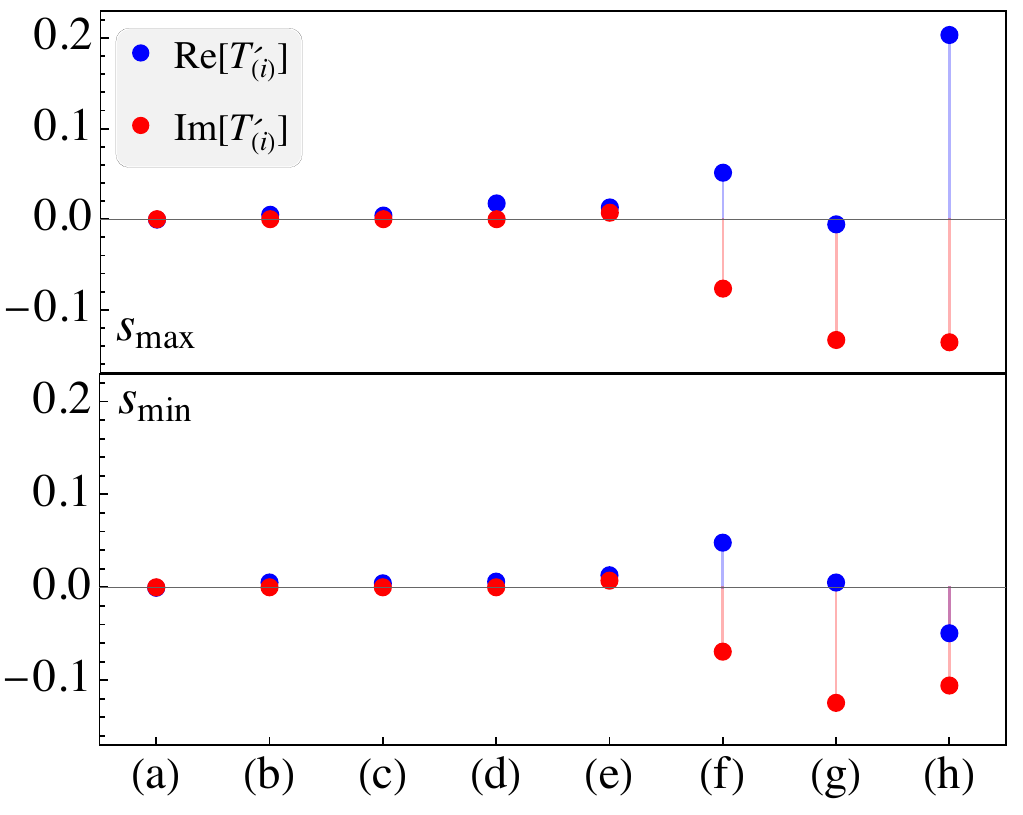} 
\caption{
Contributions to $T$ (left) and $T^\prime$ (right) from each diagram in Fig.~\ref{fig:Feyn-LR} at $M_\pi = \delta/2$ for $s=s_\mathrm{max}$ (upper) and $s=s_\mathrm{min}$ (lower). The real and imaginary parts are denoted by blue and red dots. }
\label{fig:dia-TTp}
\end{figure}
As can be seen from the right panel of Fig.~\ref{fig:T-Tprime}, the lineshapes of $T^\prime$ depend only weakly on $s$ for $M_\pi > \delta$, but strongly for $M_\pi < \delta$. This strong dependence originates from threshold cusps at $M_\pi = \delta/2$ ($s=s_{\mathrm{max}}$) and $M_\pi = \delta/\sqrt{8}$ ($s=s_{\mathrm{max}}/2$). The numerical values of $T^\prime$ are larger than those of $T$ by about one order of magnitude, in line with the power counting specified in Table \ref{tab:Ti}. The individual contributions from the loop diagrams to $T^\prime$ are analyzed as follows: 
(i) Diagrams (a) and (e) yield pion-mass-independent contributions that are negligible (see Eqs.~\eqref{eq.tffs.a} and~\eqref{eq.tffs.e}).  
(ii) The contributions from diagrams (b), (c), (f), and (g) are displayed separately in Fig.~\ref{fig:pion-mass-Tp-fg}. Among those, diagrams (b) and (c) are negligible while both (f) and (g) provide sizable contributions.  The cusp-like structures in $T^\prime_{(f)}$ are caused by the $\pi N$ threshold and a nearby triangle singularity.
(iii) Diagrams (d) and (h), shown in Fig.~\ref{fig:pion-mass-Tp-dh}, exhibit similar pion-mass dependence. However, the contribution from diagram (h) is approximately an order of magnitude larger. Both diagrams feature threshold cusps at $M_\pi = \delta/2$ (for $s=s_{\mathrm{max}}$) and at $M_\pi = \delta/\sqrt{8}$ (for $s=s_{\mathrm{max}}/2$). Diagram (g) additionally produces a cusp-like structure at $M_\pi = \delta$, resulting from threshold and a proximate triangle singularity.

The above analysis shows that the contributions to $T$ and $T^\prime$ from diagrams (a)--(d), which contain a $\Delta$ propagator, remain much smaller than the total amplitude over most of the pion-mass range considered, in particular for small pion masses. Although these diagrams are not suppressed by power counting, their smaller numerical impact is consistent with an additional suppression associated with the $\Delta$-nucleon mass splitting $\delta$ when $\delta$ is larger than the pion mass~\cite{Pascalutsa:2002pi}.

For a concrete comparison near the physical point, Fig.~\ref{fig:dia-TTp} shows the evaluated contributions at $M_\pi = \delta/2 = 0.136$ GeV. All diagrams are included, with the following exceptions: contributions to $T$ from diagrams (b), (d), (g), and (h) vanish identically and are therefore not plotted.

\subsection{Threshold cusp and triangle singularity}
\label{subsec:Singularity}
In this subsection, we discuss in more detail the threshold cusp and triangle singularity that appear in Feynman diagrams (d), (f) and (h) in Fig.~\ref{fig:Feyn-LR}.
As the two simplest classes of Landau singularities of loop diagrams, their locations are determined entirely by particle kinematics, such as masses and energies, regardless of the reaction dynamics. These phenomena originate from processes for which each individual vertex of a loop diagram can be viewed as a classical process (i.e., intermediate particles go on-shell and satisfy further kinematical conditions)~\cite{Coleman:1969sm}, and are systematically described by the Landau equations \cite{Landau:1959fi}. 
The threshold cusp arises from a square-root branch point at a two-body threshold and produces a cusp structure in an $S$-wave amplitude.
The triangle singularity arises when three intermediate particles in a loop simultaneously satisfy on-shell conditions and are collinear, producing a logarithmic singularity in the amplitude. For further details, see Refs.~\cite{Bayar:2016ftu,Guo:2019twa}.

In diagrams (d) and (h), the two intermediate pions in the loops can go on-shell simultaneously when $\sqrt{s} \geq 2M_\pi$. 
This produces a cusp in the vicinity of $M_\pi = \sqrt{s}/2$, as shown in Fig.~\ref{fig:pion-mass-Tp-dh}. 
For diagram (d), it also gives rise to the imaginary part of $T^\prime$ for $M_\pi < \sqrt{s}/2$. 
For diagram (h), it significantly enhances both the existing real and imaginary parts.

\begin{figure}[t]
\includegraphics*[width=0.6\linewidth]{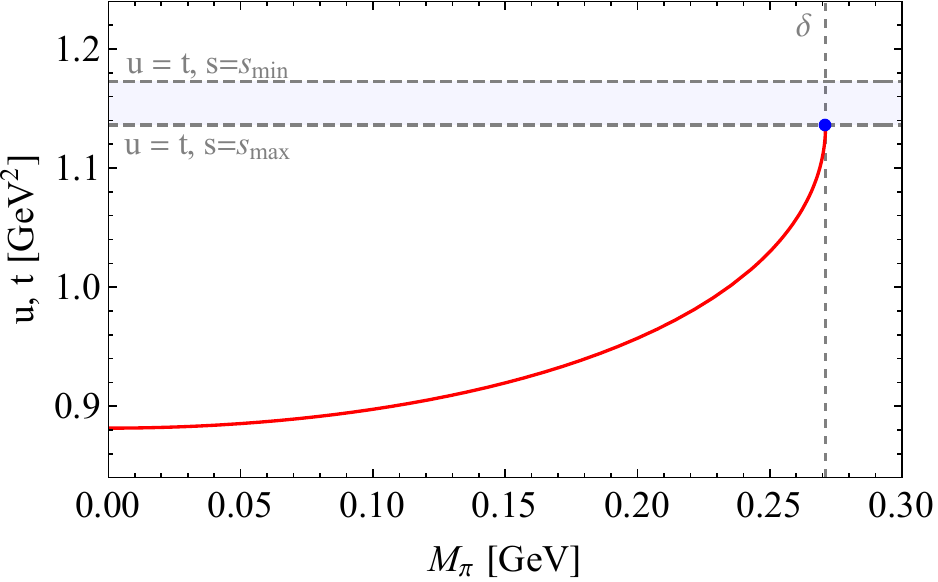}  
\caption{
Pion mass dependence of the triangle-singularity condition for $t$ and $u$ (red line) arising from Feynman diagrams (f) and (h). The vertical gray dashed line is the corresponding upper bound of $M_\pi$. The light-blue band represents the allowed phase-space region for $u=t$ bounded by $s_\mathrm{max}$ and $s_\mathrm{min}$. The blue point represents the only triangle-singularity position for $u=t$.
 } 
\label{fig:TS}
\end{figure}

In diagrams (f) and (h), the intermediate pion-nucleon pair, attached to the initial $\Delta$, becomes simultaneously on-shell if $m_\Delta \geq M_\pi + m_N$. This leads to a threshold cusp at $M_\pi = m_\Delta - m_N = \delta$, visible in Figs.~\ref{fig:pion-mass-ReT}, \ref{fig:pion-mass-Tp-fg} and \ref{fig:pion-mass-Tp-dh}. The cusp observed here is markedly sharper than its counterpart in Ref.~\cite{Bernard:2009mw} (see Fig.~5 therein), due to amplification by a proximate triangle singularity. Such a singularity arises when the intermediate pion, neutrino, and nucleon can be on-shell concurrently, and the neutrino moves along the same direction as the nucleon and moves faster than the nucleon in the $\Delta$ rest frame, for the following specific ranges of $M_\pi$, $t$, and $u$:~\cite{Guo:2019twa}.
\begin{subequations}
\begin{eqnarray}
  0 \leq M_\pi    & \leq & m_\Delta - m_N = \delta\ ,\\
   m_N^2 \leq t, u & \leq & m_N^2 + m_N M_\pi \leq m_\Delta m_N\ ,
  \label{eq:TS-range-tu}
\end{eqnarray}
\label{eq:TS-range}
\end{subequations}
\!\!where we have set $m_e=0, m_\nu=0$. 
As for diagram (h), which is a box diagram, triangle singularity is its subleading Landau singularity and occurs by shrinking the pion propagator attached to the final-state nucleon to a point.
The kinematic location of triangle singularity is determined by solving~\cite{Guo:2019twa,He:2021kkk}\footnote{In Eq.~(1.28) of Ref.~\cite{He:2021kkk}, the $2m_3^2$ should be $2sm_3^2$.}
\begin{equation}
    \lambda^{1/2}(m_\Delta^2, u,0)\lambda^{1/2}(m_\Delta^2,m_N^2,M_\pi^2) =(m_\Delta^2-m_N^2) (m_\Delta^2- u) - M_\pi^2 (m_\Delta^2+ u)\ ,
    \label{eq:TS-condition}
\end{equation}
where $u$ and $M_\pi$ are constrained by Eqs.~\eqref{eq:TS-range}, and $\lambda (x,y,z) = x^2 + y^2 + z^2 - 2xy - 2xz - 2yz$ is the Källén function. Solving for $M_\pi^2$ yields the explicit expression in terms of $u$:
\begin{equation}
M_\pi^2 = {(m_\Delta^2 - u)(u - m_N^2)}/{u}.
\end{equation}
The corresponding solution as a function of $t$ is obtained by replacing $u$ with $t$. Both relations are plotted against $M_\pi$ in Fig.~\ref{fig:TS}.
In the special case $t=u$, the kinematically allowed region is restricted to $m_\Delta m_N \leq t=u \leq(m_\Delta^2+m_N^2)/2$, indicated by the light blue band in Fig.~\ref{fig:TS}.
Within this band, the triangle singularity condition is satisfied only at the blue point, corresponding to $M_\pi = \delta$ and $t = u = m_\Delta m_N$; this point matches the blue point shown in Fig.~\ref{fig:Dalitz}. Consequently, a logarithmic divergence appears at $M_\pi = \delta$ for $s = s_\mathrm{max}$, whereas only finite peaks occur for $s = s_\mathrm{max}/2$ and $s = s_\mathrm{min}$ in Fig.~\ref{fig:T-Tprime}.

\subsection{Real decay amplitude in the limit $m_\Delta \to m_N$}
\label{subsec:Real-amp}

In diagrams (e)-(h), the intermediate nucleon and neutrino can always be simultaneously on-shell, resulting in complex decay amplitudes as shown in Fig.~\ref{fig:T-Tprime}. Lattice calculations, however, only access real amplitudes. To produce predictions that can be compared with future lattice studies, we therefore consider the limit $m_\Delta \to m_N$, where the amplitudes become purely real. When $m_\Delta = m_N$, the allowed three-body phase space in Fig.~\ref{fig:Dalitz} shrinks to a single point at $u=t=m_N^2$. 

The decay amplitude defined in Eq.~\eqref{eq:def-Amp} can be recast as
\begin{equation}
    \mathcal{M}_{t=u} = G \big[\bar u_N(p_2) q^\mu(T + T^\prime \gamma_5) u_\mu (p_1)\big]\big[ \bar u_L(q)C \bar u_L^T(q)\big],
\end{equation}
where the two final electrons have the same momentum $q$. 
In addition, as mentioned in Section~\ref{subsec:PCB}, the presence of $\gamma_5$ between the baryon spinors $\bar u_N(p_2)$ and $u_\mu(p_1)$ leads to cancellation of the hard scales $m_\Delta$ and $m_N$. 
In the limit of $m_\Delta = m_N$, the bilinear $\bar u_N(p_2) \gamma_5 u_\mu(p_1)$ is now identically zero.
The decay amplitude can be further reduced to 
\begin{equation}
    \mathcal{M}_{t=u} = G \big[ \bar u_N(p_2) q^\mu T u_\mu (p_1)\big]\big[ \bar u_L(q)C \bar u_L^T(q)\big]\ ,
\end{equation}
where the TFF $T$ can be decomposed into short- and long-range parts as 
\begin{align}
    T= T_\mathrm{SR}+ T_\mathrm{LR}\ .
\end{align}
In fact, $T_\mathrm{SR}$ and $T_\mathrm{LR}$ are simply the counterterm and one-loop contributions, respectively. Their explicit expressions are given by
\begin{subequations}
    \begin{align}
   T_\mathrm{SR} &=  4\sqrt{2} i h_1^R\ , \label{eq:T-SD-DelN}\\
   T_\mathrm{LR} &= 
   \frac{1}{288\sqrt{2}\pi^2 m_N^6} \bigg[
   \left(10 M_\pi^4 m_N^2-30 M_\pi^2 m_N^4-M_\pi^6+18 m_N^6\right) \log \frac{M_\pi^2}{m_N^2} 
   \notag\\
    &\hspace{1.5cm}  + 6m_N^4 \left(2 M_\pi^2+3 m_N^2\right) \log \frac{\mu^2}{M_\pi^2} 
    + m_N^2\left(3 m_N^2-2 M_\pi^2\right)\left(12m_N^2-M_\pi^2\right)  \notag \\ 
    &\hspace{1.5cm} - 2M_\pi(4m_N^2-M_\pi^2)^{5/2}\log\frac{\sqrt{M_\pi^2-4m_N^2}+M_\pi}{2m_N} 
   \bigg]\ ,  
\label{eq:T-LD-DelN}
\end{align}
\label{eq:T-LD-DelN-all}
\end{subequations}
where $\mu$ denotes the renormalization scale, which is typically chosen to be of order $m_N$. 
In the chiral limit $M_\pi\to 0$, the logarithmic divergences in the first and second lines of Eq.~\eqref{eq:T-LD-DelN} cancel with each other, leading to the finite TFF
\begin{equation}
     T = 4\sqrt{2} ih_1^R  +  \frac{1}{16\sqrt{2}\pi^2} 
        \Big(\log \frac{\mu^2}{m_N^2} + 2  \Big)\ .
\end{equation}

\begin{figure}[t]
\includegraphics*[width=0.55\linewidth]{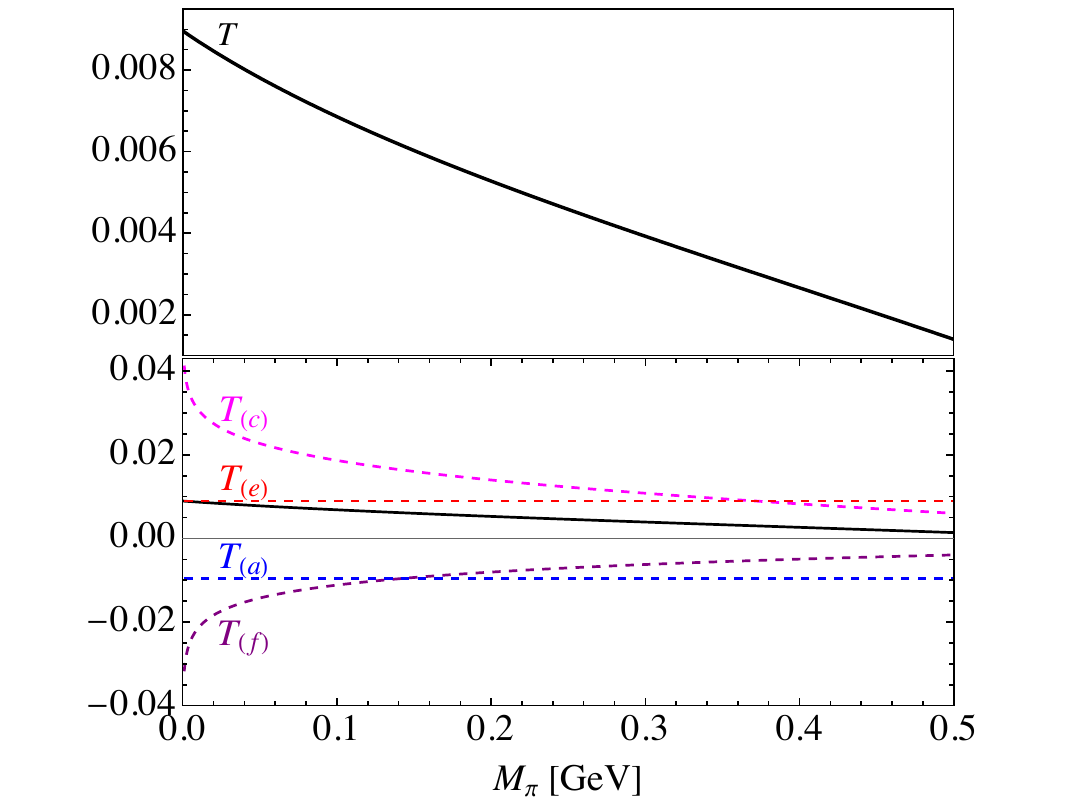}  
\caption{
Pion mass dependence of the long-range contribution to $T$ (black solid line) and the individual contributions from diagrams (a) (blue dashed line), (c) (magenta dashed line), (e) (red dashed line), and (f) (purple dashed line) with $u=t=m_N^2$ in the degenerate limit $m_\Delta = m_N$. } 
\label{fig:pion-mass-T-Delta=N}
\end{figure}

The pion mass dependence of $T_\mathrm{LR}$ is displayed in Fig.~\ref{fig:pion-mass-T-Delta=N}, plotted alongside the individual contributions from diagrams (a), (c), (e), and (f).
Marked cancellations among these contributions result in a net $T_\mathrm{LR}$ that is smaller in magnitude than any single one.
Near the physical pion mass, $M_\pi = \delta/2$, all the four contributions are of comparable size. Contributions from diagrams (a) and (c) are strongly enhanced owing to the $\Delta$-nucleon mass degeneracy, while $T_{(c)}$ and $T_{(f)}$ exhibit logarithmic divergences at $M_\pi = 0$.

\section{Summary}
\label{sec:sum}
Neutrinoless double-beta decay provides a unique probe of the Majorana nature of neutrinos and constraining the effective neutrino mass, thereby testing LNV beyond the SM. Numerous experiments have been dedicated to the search for nuclear $0\nu\beta\beta$ decays across multiple isotopes, where the elementary subprocess $nn\to ppe^-e^-$ plays a central role. The transition potential for $nn\to ppe^-e^-$ has been derived in $\Delta$-less $\chi$EFT up to NNLO in previous works, whereas a $\Delta$-full calculation remains unavailable. 
Owing to its strong coupling to the $\pi N$ system, the exchange of an intermediate $\Delta$ resonance, as illustrated in Fig.~\ref{fig:Feyn-nn-Delta-ppee}, is expected to contribute sizeably to this process. Furthermore, the $\Delta$-exchange contribution can be further enhanced by threshold cusps and triangle singularities. In this work, we initiate a $\Delta$-full treatment of $nn\to ppe^-e^-$ by studying the $0\nu\beta\beta$ decay of $\Delta^-\to p e^-e^-$. 
Such a development is essential for determining the contribution of the $\Delta$ resonance to the $nn\to ppe^-e^-$ transition matrix element, a key component in the theoretical analysis of nuclear $0\nu\beta\beta$ decays. 

We have calculated the decay amplitude of $\Delta^- \to pe^-e^-$ at one-loop level, i.e., $\mathcal{O}(p^3)$, in the framework of relativistic $\chi$EFT. The symmetric hadronic part of the decay amplitude is decomposed into a Lorentz-invariant basis with eight structure functions. We renormalize the amplitude within dimensional regularization using the $\overline{\rm MS}-1$ subtraction scheme. The renormalized amplitude contains both long-range and short-range contributions.
The long-range contribution arises from the exchange of light Majorana neutrinos through the low-energy realization of Weinberg's dimension-five operator. The short-range contribution is encoded in counterterms, which also serve canceling the loop-induced UV divergences. We construct all necessary counterterm operators and remove redundant structures using the EOMs. We find that the UV divergences in loops can indeed be exactly canceled by the counterterms.

For the kinematic configuration $t=u$, where the final two electrons are collinear, effectively behaving as a single composite particle, the eight hadronic structure functions reduce to two TFFs, $T$ and $T^\prime$.
We find that, $T$ is independent of any LECs, while $T^\prime$ depends on the axial coupling $g_A$ and the coupling $g_1$. 
The pion mass dependence of the TFFs at three representative kinematic points, $s=s_\mathrm{max}=\delta^2$, $s=s_\mathrm{max}/2$ and $s=s_\mathrm{min}=0$, are explored.
There are significant enhancements caused by the threshold cusps and triangle singularities from diagrams (d), (f) and (h). 
A diagram-by-diagram analysis further shows that, in the symmetric kinematics $t=u$, the contributions with an explicit $\Delta$ propagator are numerically subleading over most of the pion-mass range considered.

Finally, we addressed the implications for Lattice QCD. 
For large unphysical pion masses ($M_\pi > m_\Delta - m_N$), the $\Delta$ becomes stable against strong decays, making the calculation of $\Delta^-\to pe^-e^-$ feasible on the lattice. 
However, an imaginary part is present in the decay amplitude at the physical $\Delta$ mass for any pion mass, because the neutrino and nucleon in the loops can go on-shell simultaneously. We therefore investigated the decay amplitude in the degenerate $\Delta$-nucleon mass limit $m_\Delta \to m_N$, where only the TFF $T$ contributes. The explicit expressions of the short- and long-range contributions are obtained in Eqs.~\eqref{eq:T-LD-DelN-all}. 
The short-range part depends on the overall factor $G$ and the renormalized LEC $h^R_1$, which remains to be determined by future lattice simulations. In contrast, the long-range contribution is predicted entirely in terms of $G$ and the known baryon and pion masses, offering a testable benchmark for future lattice calculations.

\acknowledgments
We would like to thank Dong-Liang~Fang, Luchang Jin, Haobo Yan and Zhaolong Zhang for helpful discussions. LPH appreciates the hospitality of the Institute of Theoretical Physics (ITP) at Chinese Academy of Sciences (CAS) and Lanzhou University, where part of this work was done. This work is supported in part by the National Nature Science Foundations of China (NSFC) under Grant Nos. 12275076, 12335002, 12505105, 12125507 and 12447101; by the Fundamental Research Funds for the Central Universities under Contract No. 531118010379; and by CAS under Grant No. YSBR-101, the CAS President's International Fellowship Initiative (PIFI) (Grant No.~2025PD0022),
by the Deutsche Forschungsgemeinschaft (DFG, German Research Foundation) under Germany's
Excellence Strategy – EXC 3107 – Project-ID~533766364
and by the European Research Council (ERC) under the European Union's Horizon 2020 research and innovation programme (EXOTIC, grant agreement No. 101018170).

\appendix
\section{Interaction vertices}
\label{App:Lagrangians}


In this appendix, we give the interaction vertices from the chiral effective Lagrangians, which are needed in our calculations. 
The interaction among pions, electrons, and electron neutrinos at LO is determined by the pion decay constant in the chiral limit $F$, the Fermi coupling constant $G_F$, and the dimensionless CKM matrix element $V_{ud}$. The Lagrangian for $\pi^-\to e^-_L \bar \nu_L$ is
\begin{equation}
      \mathcal{L}_{\pi\pi}^{(2)}\supset \mathcal{L}_{\pi\to e\nu} =  2 F G_F V_{ud} \partial_\mu \pi^- \bar e_L\gamma^\mu\nu_L~.
\end{equation}
The interaction for nucleons and pions at LO is determined by $F$ and the dimensionless axial coupling constant $g_A$. The Lagrangian for $n\to p \pi^-$ is
\begin{equation}
      \mathcal{L}_{\pi N}^{(1)}\supset  \mathcal{L}_{n\to p\pi} = \frac{g_A}{\sqrt{2}F} \partial_\mu \pi^+ \bar p\gamma^\mu\gamma_5 n~.
\end{equation}
The interaction among nucleons, electrons, and electron neutrinos at LO is determined by $G_F$, $V_{ud}$, and $g_A$. The Lagrangian for $n\to p e^-_L \bar \nu_L$ is
\begin{equation}
  \mathcal{L}_{\pi N}^{(1)}\supset   \mathcal{L}_{n\to p e\nu} = 
    -\sqrt{2} G_F V_{ud} \bar e_L\gamma_\mu\nu_L
        \bar p\gamma^\mu (1-g_A\gamma_5) n\ .
\end{equation}
The Lagrangian at LO for the interaction of $\Delta$ isobars, pions, and leptons is given in Eq.~\eqref{eq:Lag-piDelta}. The interaction for $\Delta$ isobars and pions is determined by $F$ and three dimensionless coupling constants $g_1$, $g_2$, and $g_3$. 
The Lagrangian for $\Delta^- \to \Delta^0 \pi^-$ is
\begin{align} 
   \mathcal{L}_{\pi \Delta}^{(1)}\supset  \mathcal{L}_{\Delta^- \to \Delta^0 \pi} = 
     &-\frac{\sqrt{2}}{\sqrt{3}F} \partial_\alpha \pi^+
     \bar\Delta^0_\mu\bigg[ \frac{g_1}{2}\gamma^\alpha  \gamma_5g^{\mu\nu} \notag\\
     &+ \frac{g_2}{2}\left(\gamma^\mu g^{\nu\alpha} +\gamma^\nu g^{\mu\alpha}\right)\gamma_5
     +\frac{g_3}{2}\gamma^\mu\gamma^\alpha \gamma_5\gamma^\nu \bigg]\Delta^-_\nu\ . 
\end{align} 
The interaction for $\Delta$ isobars, electrons, and electron neutrinos is determined by $G_F$, $V_{ud}$, $g_1$, $g_2$, $g_3$, and the arbitrary parameter $A$. 
The Lagrangian for $ \Delta^- \to \Delta^0 e_L^-\bar\nu_L$ is
\begin{equation}
  \!\begin{aligned}
  &\mathcal{L}_{\pi \Delta}^{(1)}\supset \mathcal{L}_{\Delta^- \to \Delta^0 e\nu}
 = \frac{2\sqrt{2}}{\sqrt{3}} G_F V_{ud}  \bar\Delta^0_\mu
 \bigg\{ \frac32 \Big[ \gamma_\alpha g_{\mu\nu}
       +A (\gamma_\mu g_{\nu\alpha} + \gamma_\nu g_{\mu\alpha} ) 
      + \frac{1}{2}(3A^2+2A+1)\gamma_\mu \gamma_\alpha \gamma_\nu \Big]  \\
  &\hspace{3cm}
    -\frac{g_1}{2}\gamma_\alpha \gamma_5g_{\mu\nu}
   -\frac{g_2}{2}\left(\gamma_\mu g_{\nu\alpha} + \gamma_\nu g_{\mu\alpha}\right)\gamma_5
      -\frac{g_3}{2}\gamma_\mu\gamma_\alpha  \gamma_5\gamma_\nu \bigg\}\bar e_L\gamma^\alpha \nu_L  \Delta^-_\nu\ .
  \end{aligned}
\end{equation}
The Lagrangian at LO for the interaction of $\Delta$ isobars, nucleons, pions, and leptons is given in Eq.~\eqref{eq:Lag-piNDelta}, where there is one new coupling constant $h$. 
The interaction for $\Delta$ isobars, nucleons, and pions is determined by the pion decay constant $F$, $h$, and the arbitrary parameter $z$. The Lagrangian for the decay $\Delta \to N \pi^-$ is
\begin{subequations}
\begin{eqnarray}
     \mathcal{L}_{\pi N\Delta}^{(1)}\supset  \mathcal{L}_{\Delta \to N \pi} 
       =
       \frac{h}{F}  \partial_\mu \pi^+ \bar n (g^{\mu\nu} + z \gamma^\mu \gamma^\nu) \Delta^-_\nu  
       + \frac{h}{\sqrt{3}F} \partial_\mu \pi^+  \bar p (g^{\mu\nu} + z \gamma^\mu \gamma^\nu) \Delta^0_\nu\ .
\end{eqnarray}    
\end{subequations}
The interaction for $\Delta$'s, nucleons, and leptons is determined by $h$, $G_F$, $V_{ud}$, and $z$. 
The Lagrangian for $\Delta \to N e_L^-\nu_L$ is
\begin{subequations}
\begin{eqnarray}
    \mathcal{L}_{\pi N \Delta}^{(1)}\supset    \mathcal{L}_{\Delta \to N e\nu}
        &=& 
        2 hG_F V_{ud} \bar e_L\gamma_\mu \nu_L \bar n (g^{\mu\nu} + z \gamma^\mu \gamma^\nu) \Delta^-_\nu \\
        && + \frac{2 h}{\sqrt{3}}G_F V_{ud}\bar e_L\gamma_\mu \nu_L \bar p (g^{\mu\nu} + z \gamma^\mu \gamma^\nu) \Delta^0_\nu\ .
    \end{eqnarray}    
\end{subequations}
Finally, for easy reference, we also show the propagator for the $\Delta$ field, which reads 
\begin{equation}
\begin{split}
S^{\mu\nu}_{\Delta}=   &- \frac{\slashed{p}  + m_\Delta } {p^2 -m_\Delta^2+i\epsilon} \left[g^{\mu\nu} -\frac13 \gamma^\mu\gamma^\nu +\frac{1}{3m_\Delta} (p^\mu\gamma^\nu - \gamma^\mu p^\nu) -\frac{2}{3m_\Delta^2} p^\mu p^\nu \right]\\
   & +\frac{1}{3 m_{\Delta}^2} \frac{1+A}{1+2 A}\left\{\left[\frac{A}{1+2 A} m_{\Delta}-\frac{1+A}{2(1+2 A)} \slashed p\right]
    \gamma^\mu \gamma^\nu-\gamma^\mu p^\nu-\frac{A}{1+2 A} p^\mu \gamma^\nu\right\}.
\end{split}
\end{equation}
It reduces to the first line if $A=-1$. 

\section{Construction of counterterm Lagrangians}
\label{sec.counterterms}
There are three relevant terms with a single partial derivative acting on either the nucleon, the $\Delta$ isobar, or the two indistinguishable electrons. 
Because the electrons are identical, $q_1$ and $q_2$ must be symmetric in the Feynman rules. 
The only symmetric combination is proportional to $q_1^\mu+q_2^\mu$, which is equivalent to $p_1^\mu-p_2^\mu$. 
Using the EOM for the $\Delta$, it further reduces to $p_2^\mu$. 
Therefore, the independent counterterm Lagrangian with one derivative takes the form
\begin{equation}
    \mathcal{L}_{1,\mathrm{ct}} 
             =
            G\bar N(\overleftarrow{\partial}^\mu - \Gamma^\mu) (h_1+h_1^\prime\gamma_5) u^\dagger \tau^+u\tau^iu^\dagger\tau^+u\Psi_\mu^i \bar e_LC \bar e_L^T + \text{H.c.},
\label{eq:CC1}
\end{equation}
where $u^\dagger \tau^+u\tau^iu^\dagger\tau^+u$ has been inserted between the fields to reproduce the correct chiral transformation properties and $h_1$ and $h_1^{\prime}$ are dimensionless coupling constants.
We have factored out $G=(2\sqrt{2}G_F V_{ud})^2 m_{\beta\beta} h$ to get similar form as the amplitude in Eq.~\eqref{eq:def-Amp} and to facilitate the power-counting analysis.

When two partial derivatives are included in the counterterm Lagrangian, there are three possibilities:
\begin{enumerate}[label=(\alph*)]    
\item 
Both partial derivatives act on the nucleon field.
Using the EOM for the nucleon, this term can be reduced to the same form as that in Eq.~\eqref{eq:CC1} and is therefore redundant.
\item 
One partial derivative acts on the nucleon field and the other on the $\Delta$ field. 
This term is also equivalent to Eq.~\eqref{eq:CC1} and is therefore redundant. 
\item 
Both partial derivatives act on the electron fields:
\begin{equation}
    \mathcal{L}_{2,\mathrm{ct}}
             =
             G\bar N \gamma_\nu(h_2+h_2^\prime\gamma_5)u^\dagger \tau^+u\tau^iu^\dagger\tau^+u\Psi_\mu^i  (\partial^\mu\bar e_L) C (\partial^\nu\bar e_L^T)  + \text{H.c.}\ .\\
             \label{eq:LCC-2}
\end{equation}
Other possible structures, such as a term proportional to $(\partial^\mu \partial^\nu \bar e_L) C \bar e_L^T$, are equivalent to Eq.~\eqref{eq:LCC-2} upon the use of the EOMs and up to total derivatives.
\end{enumerate}

When three partial derivatives are included in the counterterm Lagrangian, there are five possibilities:
\begin{enumerate}[label=(\alph*)]
\item \label{item:CC3-1}
    All partial derivatives act on the nucleon field. 
    Applying the EOMs, this term can be reduced to the same form as that in Eq.~\eqref{eq:CC1} and is therefore redundant.
 
    \item \label{item:CC3-2}
    Two partial derivatives act on the nucleon field and the other one on the $\Delta$ field: 
    \begin{equation}
        F_\mathrm{ct,b}^{\prime\prime\prime}
        =(\partial^\mu \partial^\nu\bar N)\gamma_\rho\gamma_\nu \tau^+\tau^i\tau^+(\partial^\rho\Psi_\mu^i) \bar e_L C \bar e_L^T+ \text{H.c.}\ ,
    \label{eq:CC3B}
    \end{equation}
    where we omit the chiral connection $\Gamma_\mu$ and pion field $u$ for simplicity.
    \item  
    One partial derivative acts on the nucleon field and the other two on the $\Delta$ field. 
    This term can be reduced to the same form as that in Eq.~\eqref{eq:CC1} and is therefore redundant.
    \item   \label{item:CC3-4}
    One partial derivative acts on the nucleon field and the other two on the electron fields: 
    \begin{subequations}
        \begin{eqnarray}       F_\mathrm{ct,d(1)}^{\prime\prime\prime} &=& (\partial^\mu\bar N) \tau^+\tau^i\tau^+\Psi_\mu^i (\partial^\nu\bar e_L) C (\partial_\nu\bar e_L^T)+ \text{H.c.} ,
            \label{eq:CC3-Da}\\
     F_\mathrm{ct,d(2)}^{\prime\prime\prime} &=& (\partial_\nu\bar N) \gamma^\mu\gamma_\rho \tau^+\tau^i\tau^+ \Psi_\mu^i (\partial^\nu\bar e_L) C (\partial^\rho\bar e_L^T)+ \text{H.c.},
            \label{eq:CC3-Db}\\
        F_\mathrm{ct,d(3)}^{\prime\prime\prime} &=& (\partial^\nu\bar N) \gamma_\rho\gamma_\nu \tau^+\tau^i\tau^+ \Psi_\mu^i (\partial^\mu\bar e_L) C (\partial^\rho\bar e_L^T)+ \text{H.c.}.
        \end{eqnarray}
    \end{subequations}
    \item \label{item:CC3-5}
    One partial derivative acts on the $\Delta$ field and the other two on the electron fields: 
    \begin{subequations}
        \begin{eqnarray}
            F_\mathrm{ct,e(1)}^{\prime\prime\prime} &=& \bar N\gamma^\mu\gamma_\rho\tau^+\tau^i\tau^+ (\partial_\nu\Psi_\mu^i )(\partial^\nu\bar e_L) C (\partial^\rho\bar e_L^T)+ \text{H.c.},\\
            F_\mathrm{ct,e(2)}^{\prime\prime\prime} &=& \bar N\gamma^\nu\gamma_\rho\tau^+\tau^i\tau^+ (\partial_\nu\Psi_\mu^i )(\partial^\mu\bar e_L) C (\partial^\rho\bar e_L^T) + \text{H.c.}.
        \end{eqnarray}
    \end{subequations}
\end{enumerate}
However, these possible operators are not all linearly independent once the EOMs are applied: only two independent structures remain. 
We choose the first two forms $ F_\mathrm{ct,d(1)}^{\prime\prime\prime}, F_\mathrm{ct,d(2)}^{\prime\prime\prime}$ in scenario \ref{item:CC3-4} as independent counterterm structures for $\Delta\to p\, e^-e^-$ transition with three partial derivatives. The corresponding Lagrangians with correct transformation properties can be written as
\begin{equation}
\begin{split}
\mathcal{L}_{3,\mathrm{ct}}
             =&
             G\bar N(\overleftarrow{\partial}_\sigma - \Gamma_\sigma)\big[ g^{\sigma\mu}g_{\nu\rho}(h_{3a}+h_{3a}^\prime\gamma_5)  
             + g_\nu^\sigma\gamma^\mu\gamma_\rho (h_{3b}+h_{3b}^\prime\gamma_5)  \big] \\
             &  \hspace{0.5cm}
             \times u^\dagger \tau^+u\tau^iu^\dagger\tau^+u \Psi_\mu^i (\partial^\nu\bar e_L) C (\partial^\rho\bar e_L^T) + \text{H.c.}\ , 
             \label{eq:LCC-3}
    \end{split}
\end{equation}
where
$h_i^{(\prime)} (i=3a, 3b)$ are coupling constants with mass dimension $-2$.

\bibliography{References}
\end{document}